\documentclass[%
 reprint,
 amsmath,amssymb,
 aps,
]{revtex4-1}

\usepackage{dcolumn}
\usepackage{bm}
\usepackage{color}

\usepackage{amssymb,amsmath}
\usepackage{bm}

\usepackage[normalem]{ulem}

\usepackage{bm}
\usepackage{multirow}
\usepackage{mathrsfs}
\usepackage{graphicx}
\usepackage{epstopdf}

\usepackage[caption=false]{subfig}
\usepackage{epstopdf, epsfig}

\usepackage{float}

\begin{document}

\preprint{APS/123-QED}

\title{Grid refinement for entropic lattice Boltzmann models}

\author{B. Dorschner}
  \email{benedikt.dorschner@lav.mavt.ethz.ch}
\author{N. Frapolli}
  \email{*frapolli@lav.mavt.ethz.ch}
\author{S.S. Chikatamarla}%
 \email{chikatamarla@lav.mavt.ethz.ch}
 \author{I.V. Karlin}%
 \email{Corresponding author; karlin@lav.mavt.ethz.ch}
\affiliation{Aerothermochemistry and Combustion Systems Lab, Department of Mechanical and Process Engineering, ETH Zurich, CH-8092 Zurich, Switzerland}

\date{\today}

\begin{abstract}

We propose a novel multi-domain grid refinement technique with extensions to entropic incompressible, thermal and compressible lattice Boltzmann models.
Its validity and accuracy are accessed by comparison to available direct numerical simulation and experiment for the simulation 
of iso-thermal, thermal and viscous supersonic flow.
In particular, we investigate the advantages of grid refinement for the set-ups of turbulent channel flow, flow past a sphere, 
Rayleigh-B\'enard convection as well as the supersonic flow around an airfoil. 
Special attention is payed to analyzing the adaptive features of entropic lattice Boltzmann models for multi-grid simulations.


\end{abstract}

\maketitle


\section{\label{sec:introduction}Introduction}
 
Over the past decades the lattice Boltzmann method (LBM), with its roots in kinetic theory, 
has evolved into a mature and highly efficient approach to computational physics, successful in regimes 
ranging from incompressible turbulence, multiphase, thermal and compressible flows up to micro-flows and
relativistic hydrodynamics \cite{succi2015lattice}. In contrast to conventional numerical methods, LBM describes the 
governing equations in terms of discretized particle distribution functions (populations) $f_i(\bm{x},t)$ 
associated with a set of discrete velocities $ \bm{c}_i, i=1,\cdots,Q$. The 
latter are designed to recover the governing equations in the hydrodynamic limit and the velocity set spans a 
regularly spaced lattice, which algorithmically reduces to a highly efficient stream-collide algorithm 
with exact propagation and local non-linearity.

Despite these advantages, intrinsic stability issues of the original Bhatnagar-Gross-Krook lattice 
Boltzmann model (LBGK, see \cite{bgk1954}), limited the LBM for long to resolved, low Reynolds number flows and thus 
precluded a significant impact on the field of fluid dynamics. Among a number of suggestions \cite{dellar2001bulk,d2002multiple,latt2006lattice,zhang2006efficient}, this issue was resolved in \cite{Karlin1999}
by introducing the concept of a discrete $H-$function into the framework of LBM, which 
led to the development of the parameter-free and non-linearly stable entropic lattice Boltzmann method (ELBM),
consistent with the second law of thermodynamics.
In the context of multi-relaxation time LBM, the so-called KBC (Karlin-B\"osch-Chikatamarla) model, was recently proposed 
in \cite{Karlin2014} as an extension of ELBM with outstanding stability properties. Accuracy, robustness and 
reliability of entropy-based LBM was established for both periodic and wall-bounded turbulence in the isothermal 
regime in \cite{Chikatamarla2013,Bosch2015,Dorschner2016}. Similarly, the entropy concept 
has been successfully applied to thermal flows \cite{Ansumali2005b,Karlin2014} and
in combination with recent advances in the theory of admissible high-order lattices \cite{Chikatamarla2009,frapolli2014multispeed}
extended to the fully compressible regime \cite{Frapolli2015}.

However, the computational effort needed for the simulation of realistic engineering 
applications on a uniform mesh with reasonable spatial resolution is high.
On the other hand, in many cases, the small scale flow structures are confined in only a small region
of the computational domain. This reveals significant optimization potential by 
locally embedding refined blocks into the domain. 
Conceptually, two main approaches to grid refinement can be found in the literature. 
In the first approach, a cell-centered volumetric description is employed for which 
the population are continuous over the grid interface and exact conservation of mass and momentum across the grid border may be achieved \citep{Chen2006,Rohde2006}. However, as reported in \citep{Rohde2006}, staggering effects on the fine 
level require additional filtering.
The second approach uses a node-centered description, which requires interpolation and rescaling of the 
population at the grid interface (see, e.g., \citep{Filippova1998,Dupuis2003,Tolke2009,Lagrava2012}). 
Below, the discussion is restricted to the node-centered approach.
Note that the accuracy and stability of the entire simulation relies crucially on the two-way coupling 
between fine and coarse level grids.

{In this paper, we aim to 
(i): Increase the accuracy of existing grid refinement schemes by proposing a novel coupling technique, 
which avoids the low-order time interpolation commonly used in other approaches. 
(ii): Extend the refinement methodology to the entropic thermal and compressible models through a consistent rescaling of the populations. 
(iii): Study the role the entropic stabilizers by analyzing their behavior in multi-grid simulations within and across refinement patches.}


%

The outline of this paper is as follows: In Sec.~\ref{sec:LB}, we briefly review the incompressible, thermal and compressible ELB models, 
followed in Sec.~\ref{sec:gr} by a presentation of the grid refinement algorithm and the implementation details 
for each model. 
In Sec.~\ref{sec:Results}, accuracy and range of applicability is studied for various simulations. 
In the isothermal regime, the turbulent channel flow at Reynolds numbers of ${\rm Re}_{\tau}=\{180, 590\}$ is discussed
followed by the flow past a sphere at Reynolds number ${\rm Re}=3700$.
The thermal regime is validated using the Rayleigh-B\'enard convection for Rayleigh number $\rm Ra= 10^7$ and the flow past a heated 
sphere at $\rm Re = 3700$.
Finally, in the compressible regime, we focus on the viscous supersonic flow around the NACA0012 airfoil for a 
free-stream Mach number $\rm Ma_\infty=1.5$ and Reynolds number of $\rm Re=10000$. 
While for all simulations the flow properties are compared to available direct numerical simulations or experiments, we pay spacial attention to demonstrating the adaptive features of the entropic LBM during multi-grid computations. Finally, conclusions are drawn in section V.

\section{Entropic Lattice Boltzmann Models}
\label{sec:LB}
The general lattice Boltzmann equation for the population $f_i(\boldsymbol{x},t)$ is given by
\begin{equation}
\label{eq:f_equations}
	f_i(\bm{x+c_i}, t+1)=f_i^{\prime} = (1-\beta)f_i(\bm{x},t) + \beta f_i^{\text{mirr}}(\bm{x},t),
\end{equation}
where the streaming step is given by the left-hand side and the post-collision state $f^\prime_i$ is represented 
on the right-hand side by a convex-linear combination of $f_i(\bm{x},t)$ and the mirror state 
$f_i^{\text{mirr}}(\bm{x},t)$.
The LBGK model is recovered for the mirror state specified as 
\begin{equation}
	f_i^{\text{mirr}}(\bm{x},t)= 2 f_i^{eq}  - f_i.
	\label{eq:fmirr}
\end{equation}
For the construction of the equilibrium $f_i^{eq}$, the entropy function $H$ is chosen as
\begin{equation}
	H(f) = \sum_i f_i \ln \left( \frac{f_i}{W_i} \right) , \\
	\label{eq:feq_min}
\end{equation}
and is minimized subject to the local conservation laws for mass, momentum and energy
\begin{equation}
	\sum_i \lbrace 1, \bm{c_i}, c_i^2 \rbrace f_i = \lbrace \rho, \rho \bm{u}, 2\rho E \rbrace.
	\label{eq:conservationLaws}
\end{equation}
Note that the energy equation is usually neglected for the isothermal case and the weights $W_i$ become lattice-specific constants.

\subsection{\label{sec:Isothermal}KBC model for isothermal flows}
In the isothermal case, we use the entropic multi-relaxation time (KBC) model on the $D3Q27$ lattice, 
which recovers the incompressible Navier-Stokes equation with the kinematic shear viscosity 
\begin{equation}
	\nu=c_s^2 \left( \frac{1}{2\beta}- \frac{1}{2}\right),
	\label{eq:visc}
\end{equation}
where $c_s=1/\sqrt{3}$ is the speed of sound. 
Employing the minimization procedure of the $H$-function in this case yields a simple analytical expression as pointed out 
in \citep{Ansumali2003a}:
\begin{equation}\label{eq:feq_iso}
	f_i^{eq}= \rho W_i \prod_{\alpha=1}^{D=3} \left(  2- \sqrt{1+3u_\alpha^2}  \right) \left( \frac{2u_\alpha + \sqrt{1+3u_\alpha^2}}{1-u_\alpha}  \right)^{c_{i\alpha}}, 	
\end{equation}
where the lattice weights are given as $W_i=\prod_{\alpha}^{D=3} W_{c_{i\alpha}} $ with $W_0=2/3$ and $W_{-1}=W_{1}=1/6$.

Multi-relaxation time models are based on the observation that the kinetic system typically features a 
higher dimensionality than what is strictly needed to recover the hydrodynamic equations. 
Those higher-order moments are then relaxed independently to increase stability without influencing the macroscopic dynamics.
In contrast to other MRT models, the KBC suggests that the rate at which those moments should be relaxed is determined
by maximizing the entropy.
{The derivation of KBC models was already discussed in our previous publications \citep{Karlin2014, Bosch2015, Dorschner2016}; here we 
remind the main steps.}
Let us recall the set of natural moments of the $D3Q27$ lattice given by
\begin{equation}\label{eq:moments}
	\rho M_{pqr} = \sum_i f_i c_{ix}^p c_{iy}^q c_{iz}^r \quad p,q,r \in \left\lbrace 0,1,2 \right\rbrace,
\end{equation}
where the first $D+1$ moments denote the conservation laws and the pressure tensor $\bm{P}$ is given as the second-order moment.
This moment representation, Eq.~(\ref{eq:moments}), spans a basis in which the populations can equivalently be expressed.
%
%
The populations $f_i$ may then be decomposed into three part as
\begin{equation}
	f_i = k_i + s_i + h_i,
\label{eq:popSplit}
\end{equation}
corresponding to the kinematic part $k_i$, the shear part $s_i$  and the remaining higher-order moments $h_i$.
While the kinematic part $k_i$ depends only on the conserved quantities, the shear part $s_i$ contains 
deviatoric stress tensor $\bold{P^\prime}=\bold{P}-D^{-1}  Tr(\bold{P})\bm{I}$. 
The remaining moments are contained in $h_i$. 
The present model is fully described in \citep{Dorschner2016}, while various moment representations and partitions, Eq.~(\ref{eq:popSplit}), were
discussed in \citep{Bosch2015}.
By employing this decomposition, the mirror state can be expressed as
\begin{equation}
\label{eq:f_mirr}
	f_i^{\text{mirr}} = k_i + \left( 2 s_i^{eq} -s_i \right) + \left( \left(1 - \gamma\right)  h_i + \gamma h_i^{eq}\right),
\end{equation}
where $s_i^{eq}$ and $h_i^{eq}$ indicate $s_i$ and $h_i$ evaluated at equilibrium and the parameter $\gamma$
determines the relaxation of the higher-order moments, while $\gamma = 2$ recovers LBGK.
Notice that independent of the choice of $\gamma$, the model recovers the Navier-Stokes equations with the kinematic viscosity 
as given in Eq.\ (\ref{eq:visc}).
%
%
The relaxation parameter $\gamma$ is found by minimizing the discrete $H$-function
in the post-collision state $f_i^\prime$ and can be approximated as
\begin{equation}
	\gamma = \frac{1}{\beta} - \left( 2 - \frac{1}{\beta}\right) \frac{\left< \Delta s | \Delta h\right>}{\left< \Delta h | \Delta h\right>  },
	\label{eq:gamma_min_approx}
\end{equation}
where $\Delta s_i = s_i - s_i^{\text{eq}}$ and $\Delta h_i=h_i-h_i^{\text{eq}}$ indicate the deviation from equilibrium and 
the entropic scalar product is introduced as $\left< X | Y \right> = \sum_i (X_i Y_i / f_i^{\text{eq}})$.
It has been shown in \citep{Bosch2015} that the KBC models tend to the LBGK limit, $\gamma=2$, for fully resolved simulations.

\subsection{\label{sec:Thermal} Two-population KBC model for thermal flows}

In the two-population KBC model for thermal flows, the kinetic equations, Eqs.~(\ref{eq:f_equations}), for $f$-populations are modified in order to account for a variable Prandtl number \cite{ansumali2007quasi,Karlin2013,pareschi2016thermal}. 
The second set of populations, $g$, is employed to recover the energy equation \cite{Karlin2013,pareschi2016thermal}. The lattice kinetic equations for the $f$- and $g$-populations read
\begin{equation}
\label{eq:kinetics_f_KBC_thermal}
f_i(\bm{x+c_i}, t+1) = (1-\beta)f_i+ \beta f_i^{\text{mirr}} + 2 (\beta-\beta_1) \left[ f_i^*-f^{eq}_i \right],
\end{equation}
\begin{equation}
\label{eq:kinetics_g_entr}
g_i \left(\bm{x}+\bm{c}_i,t+1 \right) =  g_i \left(\bm{x},t \right) + 2\beta \left( g^{eq}_i - g_i \right)  + 2 \left( \beta - \beta_1 \right) \left[ g_i^*- g^{eq}_i \right],
\end{equation}
where $f_i^{\text{mirr}}$ (Eq.~(\ref{eq:fmirr})) and $f^{eq}_i$ (Eq.~(\ref{eq:feq_iso})) are the same as for the isothermal model, $g^{eq}_i$ is the equilibrium of $g$-populations, $f_i^*$ and $g_i^*$ are the quasi-equilibrium states for $f$- and $g$-populations, respectively, and $\beta_1$ is a second, independent, relaxation parameter. \\
The population, $g_i^{eq}$, $f_i^*$ and $g_i^*$ are constructed using the general form
\begin{align}
	G_i ={}& W_i \left(M_0 + \frac{M_\alpha c_{i\alpha}}{T_0} \right. \nonumber \\
		& \left. + \frac{\left(M_{\alpha \beta} - M_0 T_0 \delta_{\alpha \beta} \right)\left(c_{i\alpha} c_{i\beta} - T_0  \delta_{\alpha \beta} \right)}{2T_0^2} \right).
	\label{eq:grad_general}
\end{align}
The moments to be employed for the computation of the equilibrium of $g$-populations are provided in Table \ref{tab:moments_eq_quasieq}. 
The choice of moments for the quasi-equilibrium states $f^*$ and $g^*$ depends on the Prandtl number. 
In Table \ref{tab:moments_eq_quasieq}, the moments for the quasi-equilibrium populations are provided for both regimes ($\rm Pr \le 1$ and $\rm Pr >1$) \cite{Karlin2013,pareschi2016thermal}.
\begin{table}[H]
\caption{\label{tab:moments_eq_quasieq} Moments for equilibrium and quasi-equilibrium construction.}
\begin{ruledtabular}
\begin{tabular}{c|ccc}
 $G_i$ ~~~~ & $M_0$ & $M_\alpha$ & $M_{\alpha \beta}$ \\
\colrule \vspace{-2.7mm} \\
 $g^{eq}_i$ ~~~~~~~~~~ 	& $2\rho E$ 	& $q^{eq}_{\alpha}$ 													& $R^{eq}_{\alpha \beta}$\\
 $f_i^*$, ~~ $\rm Pr \le 1$ 	& $\rho$ 		& $\rho u_{\alpha}$ 													& $P^{eq}$ \\
 $f_i^*$, ~~ $\rm Pr > 1$  	& $\rho$ 		& $\rho u_{\alpha}$ 													& $P_{\alpha \beta}$ \\
 $g_i^*$ ~~ $\rm Pr \le 1$ 	& $2\rho E$ 	& $q_{\alpha} - 2 u_\beta \left( P_{\alpha \beta} - P^{eq}_{\alpha \beta} \right)$ 		& $R^{eq}_{\alpha \beta}$ \\
 $g_i^*$ ~~ $\rm Pr > 1$  	& $2\rho E$ 	& $q^{eq}_{\alpha} + 2 u_\beta \left( P_{\alpha \beta} - P^{eq}_{\alpha \beta} \right)$ 	& $R^{eq}_{\alpha \beta}$ \\
\end{tabular}
\end{ruledtabular}
\end{table}
In Table \ref{tab:moments_eq_quasieq}, the moments are defined as follows:
%
\begin{align}
	P^{eq}_{\alpha \beta} ={}& \sum_{i=1}^{n} c_{i\alpha} c_{i\beta} f^{eq}_i = \rho T_0 \delta_{\alpha \beta } + \rho u_\alpha u_\beta,\\
	q^{eq}_{\alpha} ={}& \sum_{i=1}^{n} c_{i\alpha} g^{eq}_i = 2\rho E u_\alpha + 2\rho T_0 u_\alpha, \label{eq:qeq_expression}\\
	R^{eq}_{\alpha \beta} ={}& \sum_{i=1}^{n} c_{i\alpha} c_{i\beta}  g^{eq}_i \nonumber \\={}& 2\rho E \left( T_0 \delta_{\alpha \beta }  + u_\alpha u_\beta \right) + 2\rho T_0 \left( T_0 \delta_{\alpha \beta }  + 2 u_\alpha u_\beta \right) \label{eq:Req_expression},
\end{align}
where the total energy $E$ is defined as
\begin{equation}
E = \frac{D}{2} T + \frac{1}{2} u^2,
\end{equation}
and $D$ is the space dimension. \\
The relaxation parameters $\beta$ and $\beta_{1}$ are related to the kinematic viscosity and thermal diffusivity depending on the Prandtl number as
\[
\nu =
\begin{cases}
\left(\frac{1}{2\beta} -\frac{1}{2}\right) T_0,       & \text{if  Pr $\le$ 1,} \\
\left(\frac{1}{2\beta_1} -\frac{1}{2}\right) T_0,       & \text{if  Pr $>$ 1,}
\end{cases}
\]
\[
\alpha_{th} =
\begin{cases}
\left(\frac{1}{2\beta_1} -\frac{1}{2}\right) T_0,       & \text{if  Pr $\le$ 1,} \\
\left(\frac{1}{2\beta} -\frac{1}{2}\right) T_0,       & \text{if  Pr $>$ 1,}
\end{cases}
\]
so that the Prandtl number $\rm Pr$ reads
\[
\rm{Pr} = \frac{\nu}{\alpha_{th}} =
\begin{cases}
 \frac{(1-\beta)\beta_{1}}{(1-\beta_{1})\beta},       & \text{if  Pr $\le$ 1,} \\
 \frac{(1-\beta_1)\beta}{(1-\beta)\beta_1},       & \text{if  Pr $>$ 1.}
\end{cases}
\]

\subsection{\label{sec:Compressible} Two-population ELBM for compressible flows}

Same as for the thermal two-populations ELBM, also for the compressible model \cite{Frapolli2015,frapolli2016compressible}, we employ two populations. However, in the compressible model a multi-speed lattice, the D$d$Q$7^d$, is used and the second population is needed to change the adiabatic exponent $\gamma_{ad}$. The kinetic equations for the compressible model read
\begin{equation}
\label{eq:quasi_equilibrium_kinetic_equation_summary_f}
f_i (\bm{x}+\bm{c}_i, t+1) -f_i (\bm{x},t)= \alpha \beta_2 \left( f^{\rm eq}_i -f_i \right) +2\left(\beta_2 - \beta_3 \right) [f^{*}_i-f^{\rm eq}_i],
\end{equation} 
\begin{equation}
\label{eq:quasi_equilibrium_kinetic_equation_summary_g}
g_i (\bm{x}+\bm{c}_i, t+1) -g_i (\bm{x},t)= \alpha \beta_2 \left( g^{\rm eq}_i -g_i \right) +2\left(\beta_2 - \beta_3 \right) [g^{*}_i-g^{\rm eq}_i].
\end{equation} 
The equilibrium $f^{eq}_i$ for the $f$-populations is found by minimizing the entropy function, Eq.~(\ref{eq:feq_min}), under the constraints of conservation laws, Eq.~(\ref{eq:conservationLaws}). The minimization problem is solved with the method of the Lagrange multipliers and leads to the formal expression
\begin{equation}
\label{eq:formal_equilibrium}
f^{\rm eq}_i = \rho W_i \exp \left(  \chi + \zeta_{\alpha}c_{i\alpha} + \lambda c^2_i \right),
\end{equation}
where $\chi$, $\zeta_{\alpha}$ and $ \lambda$ are the Lagrange multipliers, which in turn are determined by solving the system of $D+2$ equations found by inserting Eq.~(\ref{eq:formal_equilibrium}) into the conservation laws Eq.~(\ref{eq:conservationLaws}). The system is solved numerically at every
node in every time-step. The equilibrium for the $g$-populations, $g_i^{\rm eq}$, can be computed directly from $f^{eq}_i$ as
\begin{equation}
\label{eq:g_equilibrium_from_f_eq}
g_i^{\rm eq}= (C_{\rm v} - D/2) T f_i^{\rm eq},
\end{equation}
where $C_{\rm v}$ is the heat capacity at constant volume. \\
The quasi-equilibrium state needs to be chosen depending on the Prandtl number \cite{frapolli2014multispeed}. For $Pr\le 1$, the quasi-equilibrium state depends on the centered heat-flux tensor $\overline{Q}_{\alpha \beta \gamma}$, which can be written in a compact form as
\begin{equation}\label{eq:QEf_summary}
f_i^*=f_i^{\rm eq} + W_i\frac{ \overline{Q}_{\alpha \beta \gamma}[c_{i\alpha} c_{i\beta} c_{i\gamma}-3Tc_{i\gamma} \delta_{\alpha \beta}]}{6 T^3},
\end{equation}
where 
\begin{equation}\label{eq:QEf_summary}
\overline{Q}_{\alpha \beta \gamma} = \sum^{Q}_{i=1}f_i (c_{i\alpha}-u_{\alpha}) (c_{i\beta}-u_{\beta}) (c_{i\gamma}-u_{\gamma}).
\end{equation}
%
The quasi-equilibrium populations $g^{*}_i$ are written consistently with the $f^{*}_i$ populations and read
\begin{equation}
\label{eq:g_star_summary}
g_i^{*} = g_i^{eq} + \frac{W_i \overline{q}_{\alpha} c_{i\alpha}}{T},
\end{equation}
where $\overline{q}_{\alpha}$ is the contracted centered heat-flux tensor associated to the internal degrees of freedom of the gas
\begin{equation}
\label{eq:contracted_q_star_summary}
\overline{q}_{\alpha} = \sum^{Q}_{i=1}g_i (c_{i\alpha}-u_{\alpha}).
\end{equation}
In the above expressions $W_i = W_i(T)$ are the temperature-dependent weights \cite{Frapolli2015,frapolli2016compressible}. \\
Finally, the relaxation parameter  $\alpha$ is computed as the positive root of the entropy condition
\begin{equation}
\label{eq:entropy_condition}
H(f+\alpha (f^{eq}-f)) = H(f),
\end{equation}
where $H$ is the entropy function (Eq.~(\ref{eq:feq_min})). 
{This model is referred here as the entropic lattice Boltzmann model (ELBM).}
The kinematic viscosity and the thermal diffusivity are thus related to the relaxation parameters $\beta_2$ and $\beta_3$ by
\begin{equation}
\label{eq:beta2}
\beta_2 = \frac{1}{\frac{2\nu}{T}+1},
\end{equation}
\begin{equation}
\label{eq:beta3}
\beta_3 = \frac{1}{\frac{2\alpha_{th}}{T}+1},
\end{equation}
and the heat capacity $C_v$ is derived from the desired adiabatic exponent $\gamma_{ad}$ from
\begin{equation}
\label{eq:C_v_from_gamma}
C_v =  \frac{1}{\gamma_{ad}-1}.
\end{equation}
The model implies a bulk viscosity $\xi$ of
\begin{equation}
\xi = \left( \frac{1}{C_v}-\frac{2}{D} \right) \mu.
\end{equation}
For further details on the model the reader is referred to \cite{Frapolli2015,frapolli2016compressible}.

\section{\label{sec:gr} Multi-domain grid refinement}

\begin{figure}
	\centering					
	\includegraphics[width=0.45\textwidth]{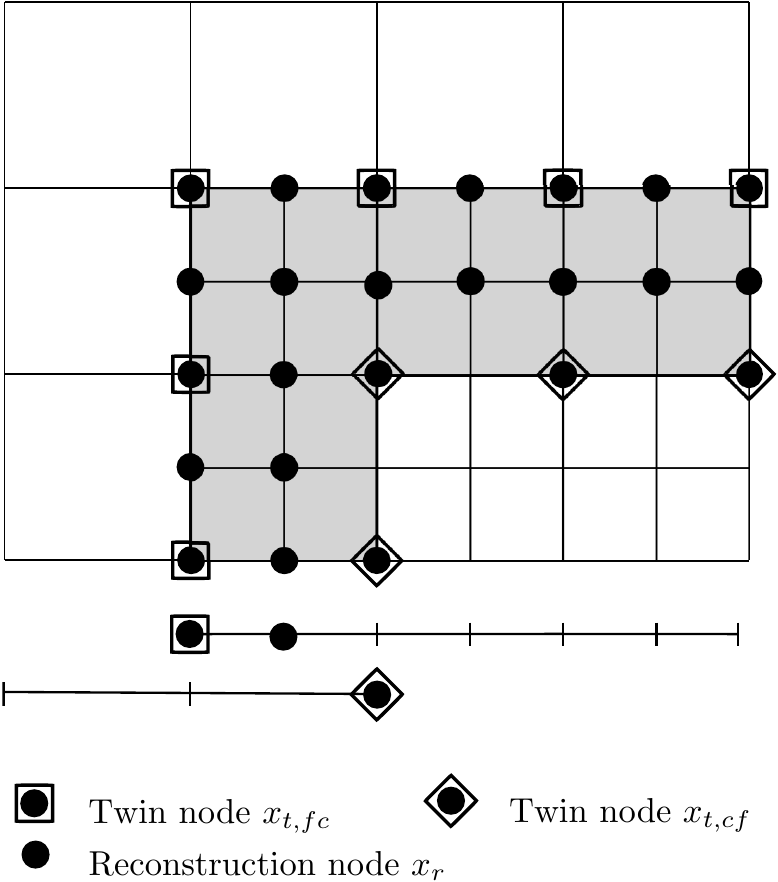}
	\caption{Schematic of the overlapping grid interface between two levels in one (bottom) and two (top) dimensions.}
	\label{fig:schematic}
\end{figure}

In the multi-domain approach, the fine-level grid patches are inserted into the coarse-level grid and only 
one level is present for each patch.
Thus, the most crucial part of the algorithm is the coupling of different grid patches, namely the information transfer
from the fine level to the coarse level and vice versa.
This two-way coupling is realized through an overlapping region which extents the nominal domain of the two patches as 
indicated by the gray shaded region in Fig.~\ref{fig:schematic}. In this region both fine and coarse grids are present.
The minimal interface width for the proposed coupling mechanism is two and one grid nodes for the fine- and coarse-level grid, respectively.
After each time step in the corresponding grid, the information at the boundary needs to be extracted from the available 
information of the neighboring patch.
It is convenient to distinguish two types of nodes, namely twin nodes $\bm{x}_t$ and reconstruction nodes $\bm{x}_r$.
Twin nodes are nodes within the interface, located in places where both coarse and fine level nodes are overlapping.
We write $\bm{x}_{t, fc}$ and $\bm{x}_{t, cf}$ to indicate twin nodes in the interface of the fine and coarse level grids, respectively.
The reconstruction nodes belong to the fine level interface and do not coincide with any coarse level nodes.
Thus the information on those nodes is not readily available and information from the neighboring grid nodes needs to be taken into account in order
to complete the information exchange between the grids.
The information extraction for the two types of nodes is conceptually different and will be elaborated in the next section.

\subsection{Twin node coupling and rescaling}
\label{sec:Rescaling}
The standard LBM is restricted to regular meshes and non-dimensional quantities are usually evaluated
with respect to the corresponding lattice units, specific to the chosen grid.
Using multiple grids with different spacing yields a set of lattice units which need to be rescaled appropriately in order to assure 
identical non-dimensional quantities. 
While the spatial scaling is straightforwardly defined as 
\begin{equation}
	r=\frac{\delta x_c}{\delta x_f},
\end{equation}
more possibilities exist for the time scale. 
While diffusive scaling relates the time and space scales as $\delta t ~\sim \delta x^2$, the 
convective scaling requires $\delta t ~\sim \delta x$. 
In this paper, we consider the convective scaling due to its greater numerical efficiency; it follows 
\begin{equation}
	r=\frac{\delta t_c}{\delta t_f}.
\end{equation}
As a result, the macroscopic quantities such as pressure, velocity and temperature are continuous across the grid border.
However, for a continuous Reynolds number $ Re=UL/\nu$ the viscosity scales as
\begin{equation}
 \nu_f=r \nu_c.
\end{equation}
On the other hand, viscosity is related to the relaxation parameter $\beta$ by Eq.\ (\ref{eq:visc}) and thus scaling of
the relaxation parameter is required as

\begin{equation}
	\beta_f= \frac{1}{1+r(1/\beta_c-1)}.
\end{equation}
For the rescaling of the populations, we decompose them into $f_i = f_i^{eq}(M_{\mathcal{C}})  + f_i^{neq}(M_{\mathcal{C}}, \nabla M_{\mathcal{C}})$, where the equilibrium distribution 
is written as a function of the conserved moments $M_\mathcal{C}$ and the non-equilibrium part additionally depends on their gradients. 
Since the conserved fields are continuous across the interface, the equilibrium does not require any rescaling.
However, the non-equilibrium part is proportional to the gradients and therefore needs rescaling.
The rescaling procedure will be discussed separately for each lattice Boltzmann model.

In the case of an external force is, we use the exact difference method as proposed in \citep{Kupershtokh2004}, 
where the post-collision populations are modified according to:
\begin{equation}
	f_i^{\prime} = f^{\prime}_i + \left[f^{eq}_{i}\left(\rho, u_{\alpha}+\delta u_{\alpha} \right) - f^{eq}_{i}\left(\rho, u_{\alpha} \right) \right],
\end{equation}
where $\delta u_{\alpha}$ is the change in velocity due to the external force $F_{\alpha}$:
\begin{equation}
	\delta u_{\alpha} = \frac{F_{\alpha}}{\rho} \delta t.
\end{equation}
%
Due to the convective time scaling employed here, the change in velocity has to be rescaled accordingly; it follows
\begin{equation}
	\delta u_f = \frac{\delta u_c}{r}.
\end{equation}

\subsubsection{Isothermal model}
In the isothermal case, we may approximate the non-equilibrium part of the population as 
\begin{equation}
f_i^{neq}  	\approx \frac{W_i}{2 c_s^4} P^{(1)}_{\alpha \beta}[c_{i\alpha} c_{i\beta}-c_s^2 \delta_{\alpha \beta}],
\end{equation}
%
where non-equilibrium part of the pressure tensor is given as 
\begin{equation}
P^{(1)}_{\alpha \beta} = \frac{c_s^2\rho}{2\beta} \left( \partial_{\alpha} u_{\beta} + \partial_{\beta} u_{\alpha} \right).
\end{equation}
Note that strain rate tensor is given in fine and coarse lattice units respectively, and therefore requires rescaling to assure continuity of the non-equilibrium fields. 
Straightforward substitution yields the relation between coarse and fine level non-equilibrium as
\begin{equation}
\label{eq:f_neq_rescaling_iso}
	f_{i,f}^{neq}= \frac{\beta_c}{r \beta_f} f_{i,c}^{neq},
\end{equation}
which identifies the scaling factor \citep{Filippova1998}.
This rescaling is applied to all twin nodes $\bm{x}_t$ (see Fig.~\ref{fig:schematic}), for the coarse and the fine grids.
{Note that the adaptive nature of entropy-based LBM does not require additional smoothing or filtering to project the fine-scale solution onto the coarse-level grid (see section \ref{sec:Results} for a more thorough discussion). This is in contrast to, e.g.,\citep{Lagrava2012}, where a box-filter is required to maintain stability.}


\subsubsection{Thermal model}

For the thermal two-population model the rescaling of the $f$- and $g$-populations cannot be applied in the same way as for the single-population 
case, Eq.~(\ref{eq:f_neq_rescaling_iso}), because two independent relaxation parameters appear in the kinetic equations. 
Moreover, we must distinguish between the two cases of $\rm Pr \le 1$ and $\rm Pr >1$. 

For $\rm Pr \le 1$ there is no contribution of the quasi-equilibrium state in the kinetic equation for $f$, so that the same rescaling as Eq.~(\ref{eq:f_neq_rescaling_iso}) can be applied for the $f$-populations. For the $g$-populations, however, this is no more valid, since the non-equilibrium part depends on two independent relaxation parameters. This can be verified by deriving an analytical expression for the non-equilibrium $g$-populations through Chapman-Enskog expansion, as in \cite{pareschi2016thermal}. The non-equilibrium part of the $g$-population depends on the higher-order, non-conserved moments as
\begin{equation}
g_{i}^{neq} = g_{i}^{neq} \left[ q^{(1)}_{\alpha}(\beta,\beta_1),R^{(1)}_{\alpha \beta}(\beta_1) \right],
\end{equation} 
where $q^{(1)}_{\alpha}$ and $R^{(1)}_{\alpha \beta}$ are the first and second order non-equilibrium moments of the $g$-populations. The $q^{(1)}_{\alpha}$ moment includes a contribution {of $\beta$, different} from the moment $R^{(1)}_{\alpha \beta}(\beta_1)$ and the higher-order moments, which need to be excluded before rescaling of the non-equilibrium part. The $q^{(1)}_{\alpha}$ moment can be written analytically as
\begin{equation}
q^{(1)}_{\alpha}(\beta,\beta_1) = -\frac{1}{2\beta_1}\rho D T_0 \partial_{\alpha}T+2u_{\beta}P^{(1)}_{\alpha \beta}(\beta),
\end{equation} 
so that the contribution in $\beta$ can be separated from the rest of the contributions to the non-equilibrium part of the $g$-populations. This contribution can be rescaled separately according to
\begin{equation}
P^{(1)}_{\alpha \beta, f} = \frac{\beta_{c}}{r \beta_{f}} P^{(1)}_{\alpha \beta, c}.
\end{equation}
After subtraction of the term dependent on $\beta$ the rescaling is performed as usual. The final result reads
\begin{equation}
g_{i,f}^{neq} =  \frac{\beta_{1,c}}{r \beta_{1,f}} \left( g_{i,c}^{neq} - W_i \frac{ 2 u_{\beta} P^{(1)}_{\alpha \beta, c} c_{i\alpha}}{T_0} \right) + W_i \frac{ 2 u_{\beta} P^{(1)}_{\alpha \beta, f}c_{i\alpha}}{T_0}.
\end{equation} 
Note that the non-equilibrium pressure tensor $P^{(1)}_{\alpha \beta}$ can be computed directly from $f$-populations by
\begin{equation}
P^{(1)}_{\alpha \beta} = \sum_i c_{i\alpha} c_{i\beta}(f_i-f^{eq}_i).
\end{equation}

For the case $\rm Pr >1$, similar considerations can be applied, and the final results are: for the $f$-populations:
\begin{equation}
\label{eq:f_neq_rescaling_thr_Prgr1}
	f_{i,f}^{neq}= \frac{\beta_{1,c}}{r \beta_{1,f}} f_{i,c}^{neq},
\end{equation}
while for the $g$-populations:
\begin{equation}
g_{i,f}^{neq} =  \frac{\beta_{c}}{r \beta_{f}} \left(g_{i,c}^{neq} - W_i \frac{ 2 u_{\beta} P^{(1)}_{\alpha \beta, c} c_{i\alpha}}{T_0} \right) + W_i \frac{ 2 u_{\beta} P^{(1)}_{\alpha \beta, f}c_{i\alpha}}{T_0},
\end{equation} 
where the non-equilibrium pressure tensor is rescaled as
\begin{equation}
P^{(1)}_{\alpha \beta, f} = \frac{\beta_{1,c}}{r \beta_{1,f}} P^{(1)}_{\alpha \beta, c}.
\end{equation}

\subsubsection{Compressible model}

As for the thermal model, also in the compressible model the populations can not be rescaled directly, since the non-equilibrium part depends on two relaxation parameters. For both $f$- and $g$-populations the non-equilibrium parts can be derived analytically as a function of the higher-order, non-conserved moments \cite{frapolli2016compressible}. We consider here only the case $\rm Pr \le 1$; for the case $Pr >1$ the relaxation parameters $\beta_2$ and $\beta_3$ needs simply to be interchanged. \\
For the $f$-populations, the non-equilibrium part depends on the higher-order, non-conserved moments as
\begin{equation}
f^{(1)}_{i} = f^{(1)}_{i} \left[ P^{(1)}_{\alpha \beta}(\beta_2), Q^{(1)}_{\alpha \beta \gamma}(\beta_2,\beta_3), R^{(1)}_{\alpha \beta \gamma \mu}(\beta_2) \right],
\end{equation}
where $P^{(1)}_{\alpha \beta}$ is the non-equilibrium pressure tensor, and $Q^{(1)}_{\alpha \beta \gamma}$ and $R^{(1)}_{\alpha \beta \gamma \mu}$ are the third- and fourth-order non-equilibrium moments. In this case, the different relaxation shows up only in the $Q^{(1)}_{\alpha \beta \gamma}$ tensor, which can be written analytically as
\begin{align}
Q^{(1)}_{\alpha \beta \gamma} (\beta_2,\beta_3) &= -\frac{1}{2\beta_3}\rho T \left[  \partial_{\alpha}T \delta_{\beta \gamma} +  \partial_{\beta}T \delta_{\alpha \gamma} +  \partial_{\gamma}T \delta_{\alpha \beta}    \right] \\ \nonumber
 &+ u_{\alpha}P^{(1)}_{\beta \gamma}(\beta_2)+ u_{\beta}P^{(1)}_{\alpha \gamma}(\beta_2)+ u_{\gamma}P^{(1)}_{\alpha \beta}(\beta_2).
\end{align}
The contribution related to $\beta_3$ can be subtracted from the rest of the non-equilibrium part and rescaled separately according to the proper relaxation.
The non-equilibrium part without the contribution related to $\beta_3$ can be written as
\begin{align}
\overline{f}^{neq}_{i} &=  f^{neq}_{i} \\ \nonumber
				& + \frac{W_iY_{i,\alpha \beta \gamma}}{6 T^3} \left( u_{\alpha}P^{(1)}_{\beta \gamma}(\beta_2)+ u_{\beta}P^{(1)}_{\alpha \gamma}(\beta_2)+ u_{\gamma}P^{(1)}_{\alpha \beta}(\beta_2) \right. \\ \nonumber
	& \left. -Q^{(1)}_{\alpha \beta \gamma}(\beta_2,\beta_3)  \right) ,
\end{align}
where
\begin{equation}
Y_{i,\alpha \beta \gamma} = c_{i\alpha} c_{i\beta} c_{i\gamma}-3 c_{i\gamma}T \delta_{\alpha \beta}.
\end{equation}
At this point the reduced non-equilibrium part can be rescaled according to
\begin{equation}
\overline{f}_{i,f}^{neq} =  \frac{\beta_{2,c}}{r \beta_{2,f}} \overline{f}_{i,c}^{neq},
\end{equation} 
{and} the final non-equilibrium populations become
\begin{align}
f_{i,f}^{neq} &=  \overline{f}_{i,f}^{neq} \\ \nonumber
				& - \frac{\beta_{3,c}}{r \beta_{3,f}} \frac{W_iY_{i,\alpha \beta \gamma}}{6 T^3} \left( u_{\alpha}P^{(1)}_{\beta \gamma,c}(\beta_2)+ u_{\beta}P^{(1)}_{\alpha \gamma,c}(\beta_2)+ u_{\gamma}P^{(1)}_{\alpha \beta,c}(\beta_2) \right. \\ \nonumber
	& \left. -Q^{(1)}_{\alpha \beta \gamma,c}(\beta_2,\beta_3)  \right) ,
\end{align}
where $P^{(1)}_{\alpha \beta}$ and $Q^{(1)}_{\alpha \beta \gamma}$ can be computed as
\begin{equation}
P^{(1)}_{\alpha \beta} = \sum_i c_{i\alpha} c_{i\beta}(f_i-f^{eq}_i),
\end{equation}
and
\begin{equation}
Q^{(1)}_{\alpha \beta \gamma} = \sum_i c_{i\alpha} c_{i\beta} c_{i\gamma} (f_i-f^{eq}_i).
\end{equation}
For $g$-populations a similar procedure is applied. The non-equilibrium part can be expressed as
\begin{equation}
g^{(1)}_{i} = g^{(1)}_{i} \left[ Tr^{(1)}(\beta_2), q^{(1)}_{\alpha}(\beta_2, \beta_3), R^{(1)}_{\alpha \beta}(\beta_2) \right],
\end{equation}
where $Tr^{(1)}$, $q^{(1)}_{\alpha}$ and $R^{(1)}_{\alpha \beta}$ are the zeroth-, first- and second-order non-equilibrium moments of $g$-populations. Similar to the $f$-populations, the different contribution on the relaxation, $\beta_3$, shows up only in the $q^{(1)}_{\alpha}$ tensor; this can be expressed analytically as \cite{frapolli2016compressible}
\begin{equation}
q^{(1)}_{\alpha} = - \frac{1}{2\beta_3} \rho T \left(  2C_v - D   \right) \partial_{\alpha} T + u_{\alpha}Tr^{(1)}(\beta_2).
\end{equation}
Also for this case then, the contribution related to $\beta_3$ can be subtracted from the rest of the non-equilibrium $g$-populations and thus can be rescaled separately according the the proper relaxation. The non-equilibrium part without the contribution related to $\beta_3$ can be written as
\begin{equation}
\overline{g}^{neq}_{i} = g^{neq}_{i} + W_i\frac{ \left( u_{\alpha}Tr^{(1)}(\beta_2)  - q^{(1)}_{\alpha}(\beta_2,\beta_3)  \right)    c_{i\alpha} }{T}.
\end{equation}
The reduced non-equilibrium part can be rescaled as
\begin{equation}
\overline{g}_{i,f}^{neq} =  \frac{\beta_{2,c}}{r \beta_{2,f}} \overline{g}_{i,c}^{neq},
\end{equation} 
thus the final non-equilibrium populations become
\begin{equation}
g_{i,f}^{neq} =  \overline{g}_{i,f}^{neq} + \frac{\beta_{3,c}}{r \beta_{3,f}} W_i \frac{\left( u_{\alpha}Tr^{(1)}(\beta_2)  - q^{(1)}_{\alpha}(\beta_2,\beta_3)  \right) c_{i\alpha} }{T},
\end{equation} 
where $Tr^{(1)}$ and $q^{(1)}_{\alpha}$ can be directly computed from populations as
\begin{equation}
Tr^{(1)} = \sum_i (g_i-g^{eq}_i),
\end{equation}
and
\begin{equation}
q^{(1)}_{\alpha} = \sum_i c_{i\alpha} (g_i-g^{eq}_i).
\end{equation}

\subsection{Reconstruction nodes}
The coupling from the coarse to the fine grid requires only the rescaling procedure as outlined above in section \ref{sec:Rescaling}.
The fine level grid however requires an additional reconstruction of the hanging or reconstruction nodes that do not correspond to a coarse level node
as indicated in Fig.~\ref{fig:schematic} by $\bm{x}_r$. 
As no information from the coarse level is available at this location, the information of the neighboring nodes is required to be included through an interpolation scheme.
Analogous to \citep{Lagrava2012}, we use a centered four point stencil in each spatial dimension of the Lagrange polynomial, which, for a generic quantity 
$\lambda$,  reads 
\begin{align}\label{eq:interp}
	\lambda(x)=&\frac{1}{16}\left( -\lambda(x-3\delta x) + 9\lambda(x-\delta x) +9\lambda(x+\delta x)) \right.  -\\ \nonumber
	 &\lambda(x+3\delta x) ).	
\end{align}
Note that in contrast to \citep{Lagrava2012}, no biase in the interpolation is introduced at the corners to avoid anisotropy effects influencing the solution.
This is particularly important for the thermal and compressible model, where a biased interpolation stencil triggers spurious artifacts at the grid interface. Further, we confirm the observation of \citep{Lagrava2012} that a second-order interpolation is not sufficient.

This interpolation is used for the macroscopic quantities of the flow field needed to compute the equilibrium part of the populations as well as the non-equilibrium part of the populations using their values at the twin nodes on the coarse level grid.

\subsection{Algorithm}
In this section, we present the proposed grid refinement algorithm. 
We aim to avoid the commonly used low-order time interpolation (see, e.g., \citep{Lagrava2012}) and 
instead replace it by a high-order spatial interpolation using Eq.~(\ref{eq:interp}).
After initialization at time $t=t_0$ we assume that all populations and macroscopic fields are specified and available everywhere on all grids.
Further a refinement ratio of $r=2$ is assumed.
Starting at $t=t_0$, the simulation is evolved by the following iterative steps:
\begin{enumerate}
\item $t=t_0+\delta_{t,f}:$ Advection and collision on the fine grid. 
								  Note that information is missing on the boundary of the fine level interface and collision should be avoided on those nodes.
\item $t=t_0+\delta_{t,c}:$ Advection on both coarse and fine level grids. 
								  Information is now missing on two boundary layers in the fine and one in the coarse grid.
\item $t=t_0+\delta_{t,c}:$ Rescaling of populations and macroscopic fields on the twin nodes $x_t$ . 
\item $t=t_0+\delta_{t,c}:$ Reconstruction of the populations on $\bm{x}_r$: Interpolation of the macroscopic quantities to compute the equilibrium part of the populations and direct interpolation of the non-equilibrium part of the populations. All information is now available again.
\item $t=t_0+\delta_{t,c}:$ Collision on all grids.
\end{enumerate}
This procedure effectively switches spatial and temporal interpolation allowing for an increase of accuracy, 
important for the simulation of a range of 
flow regimes as presented in the subsequent section.

\section{Numerical validation}
\label{sec:Results}
In this section, the proposed grid refinement technique is validated and its range of applicability is accessed in the isothermal, thermal and
compressible flow regimes.
\subsection{Isothermal flows}
In this section, we investigate accuracy and stability of the proposed grid refinement algorithm for turbulent isothermal flows using 
the examples of the flow past a sphere and the turbulent channel flow where grid refinement is crucial to obtain accurate result.
The boundary conditions used for all wall-boundary nodes is Grad's approximation as presented in \citep{Dorschner2015}.

\subsubsection{Turbulent channel flow}
\begin{table}
\caption{\label{tab:table4}%
\label{tab:channelData}
Turbulent flow in a rectangular channel. The nominal and measured Reynolds numbers are indicated by $Re_{\tau,n}$ and $Re_{\tau,m}$, respectively.}
\begin{ruledtabular}
\begin{tabular}{lccc}
Contribution						&	$Re_{\tau,n}$		&	$Re_{\tau,m}$		&	$\Delta y^+$		\\
\hline
\citet{Moser1999}				&  $180$					& 	$178.13$		&	$\sim 0.054$		\\	
present 							&  $180$					& 	$173.06$		&	$\sim  1.73$		\\	
\citet{Moser1999} 				&  $590$					& 	$587.19$		&	$\sim 0.044$		\\	
present (refined) 				&  $590$					& 	$611.87$		&	$\sim 3.06$		\\	
present (uniform)				&  $590$					& 	$613.19$		&	$\sim 12.26$		\\	
\end{tabular}
\end{ruledtabular}
\end{table}

\begin{figure}
	\centering					
	\includegraphics[width=0.45\textwidth]{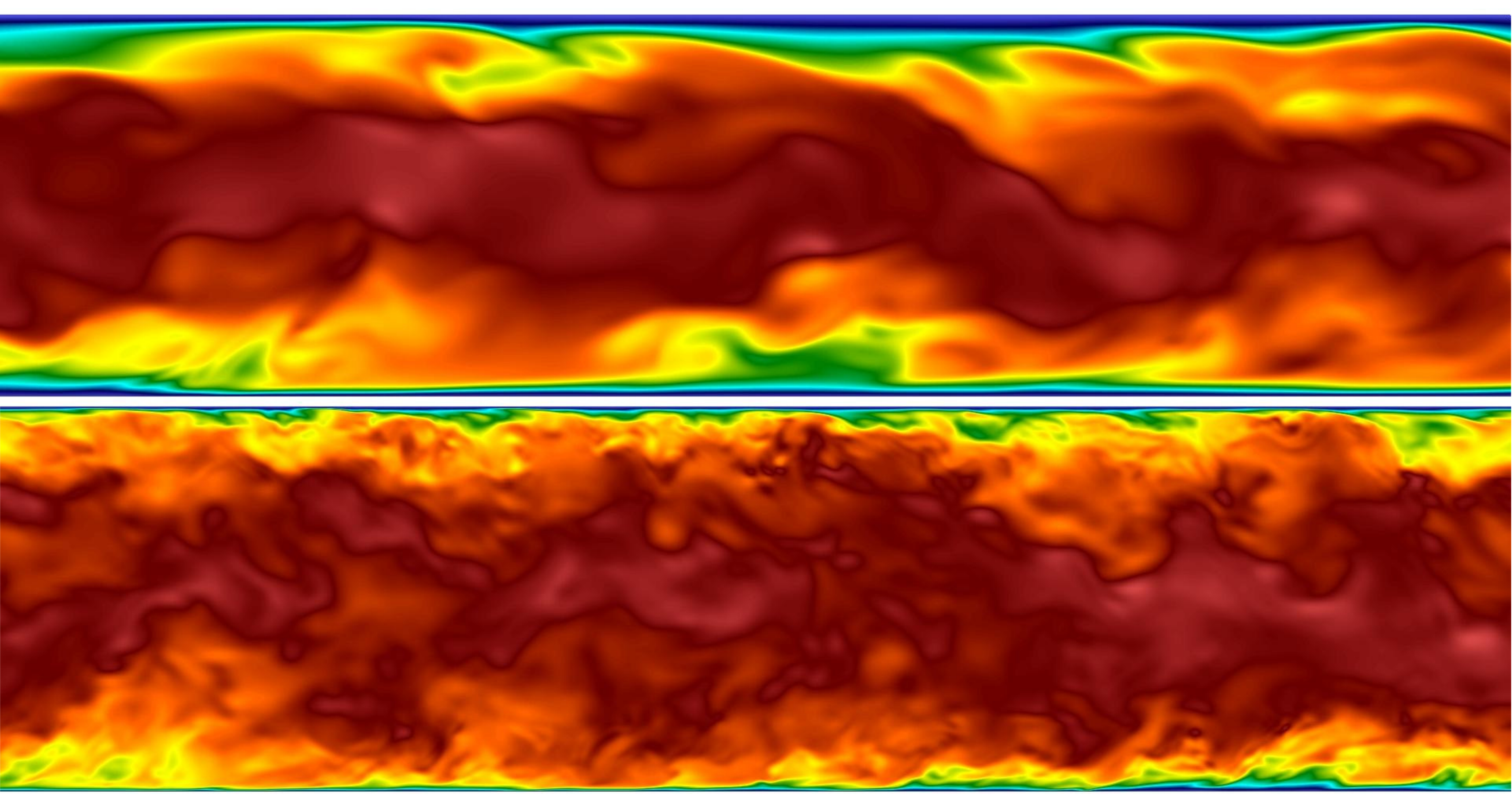}
	\caption{Slice through the turbulent channel flow for $\rm{Re}_\tau=180$ (top) and $\rm{Re}_\tau=590$ (bottom) showing a snapshot of the streamwise velocity. }
	\label{fig:channel_velSlice}
\end{figure}

\begin{figure}
	\centering					
	\includegraphics[width=0.45\textwidth]{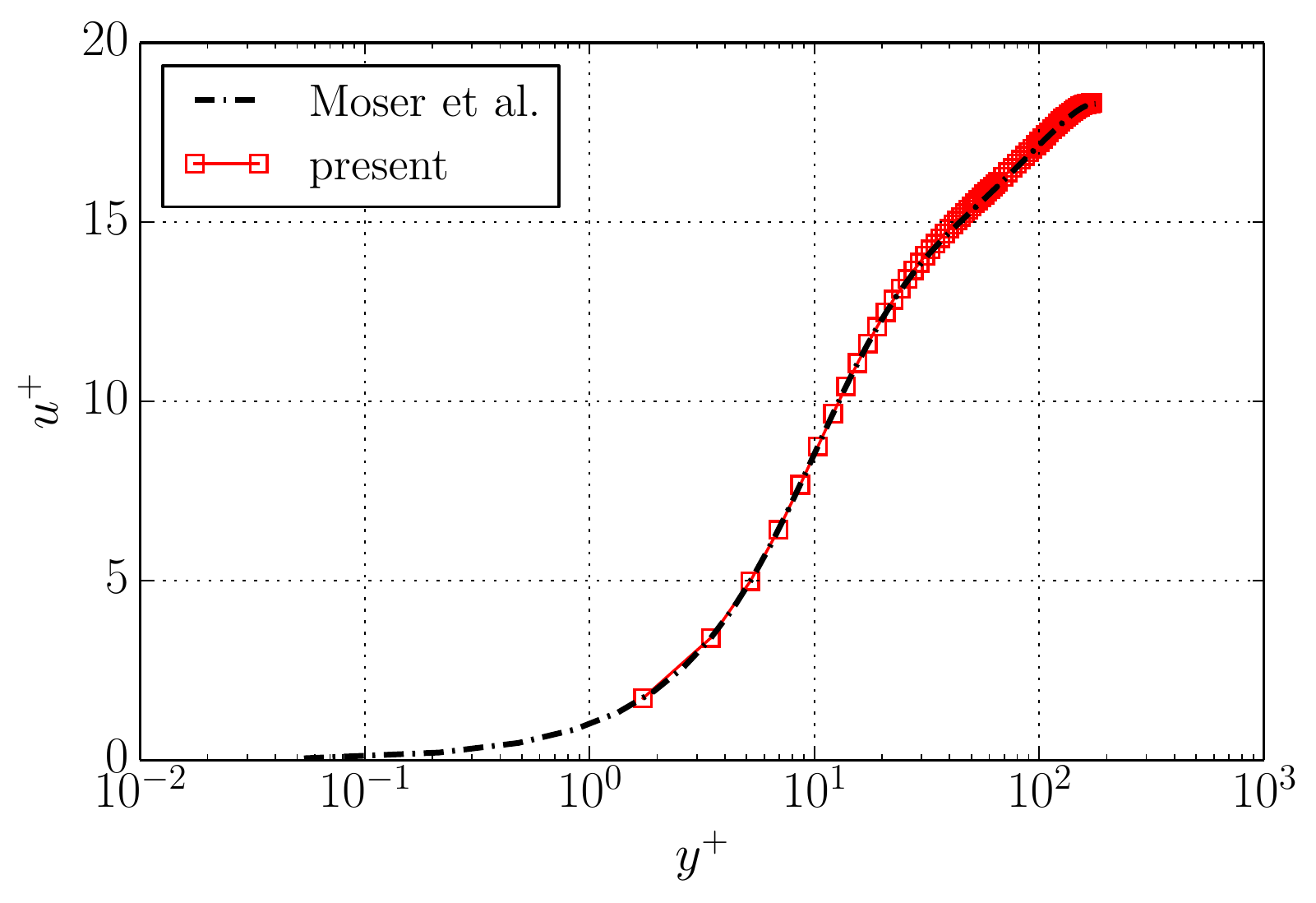}
	\caption{Mean velocity profile in a turbulent channel at $\rm{Re}_\tau=180$}
	\label{fig:channel_u_Re180}
\end{figure}
\begin{figure}
	\centering					
	\includegraphics[width=0.45\textwidth]{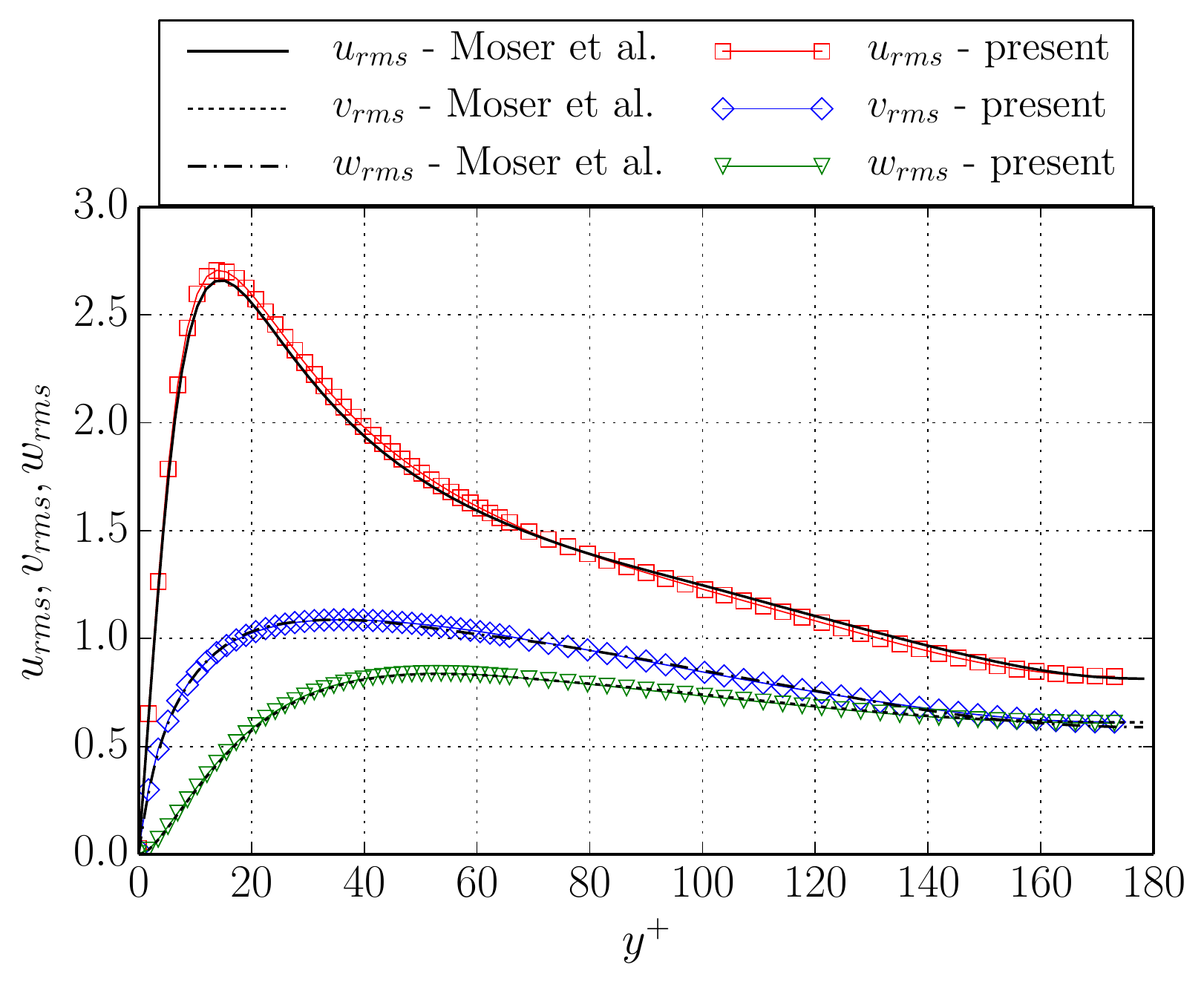}
	\caption{Rms velocity profile in a turbulent channel at $\rm{Re}_\tau=180$}
	\label{fig:channel_urms_Re180}
\end{figure}

\begin{figure*}
	\centering					
	\includegraphics[width=0.9\textwidth]{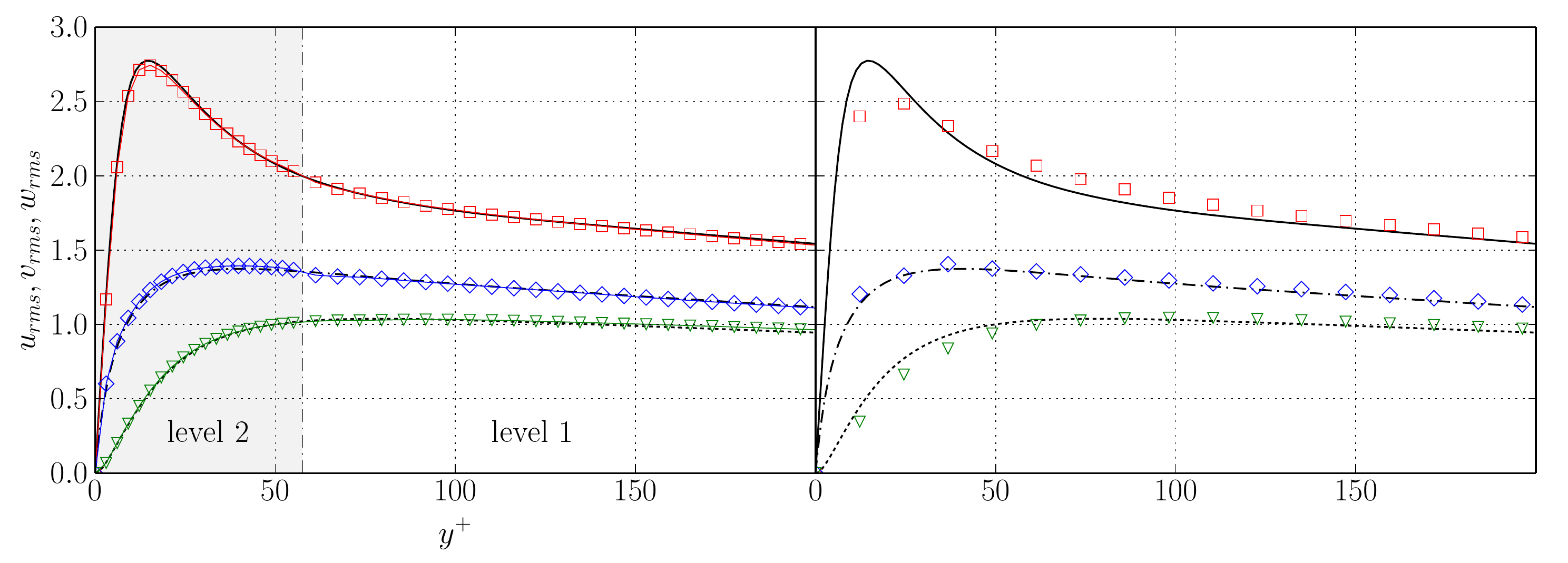}
	\caption{Rms velocity profiles in a turbulent channel at $\rm{Re}_\tau=590$ with a two-level refinement in the near-wall region (left) and 
			the non-refined case (right). For the legend please refer to Fig.~\ref{fig:channel_urms_Re180}.	  }
	\label{fig:channel_urms_Re590}
\end{figure*}

\begin{figure}
	\centering					
	\includegraphics[width=0.45\textwidth]{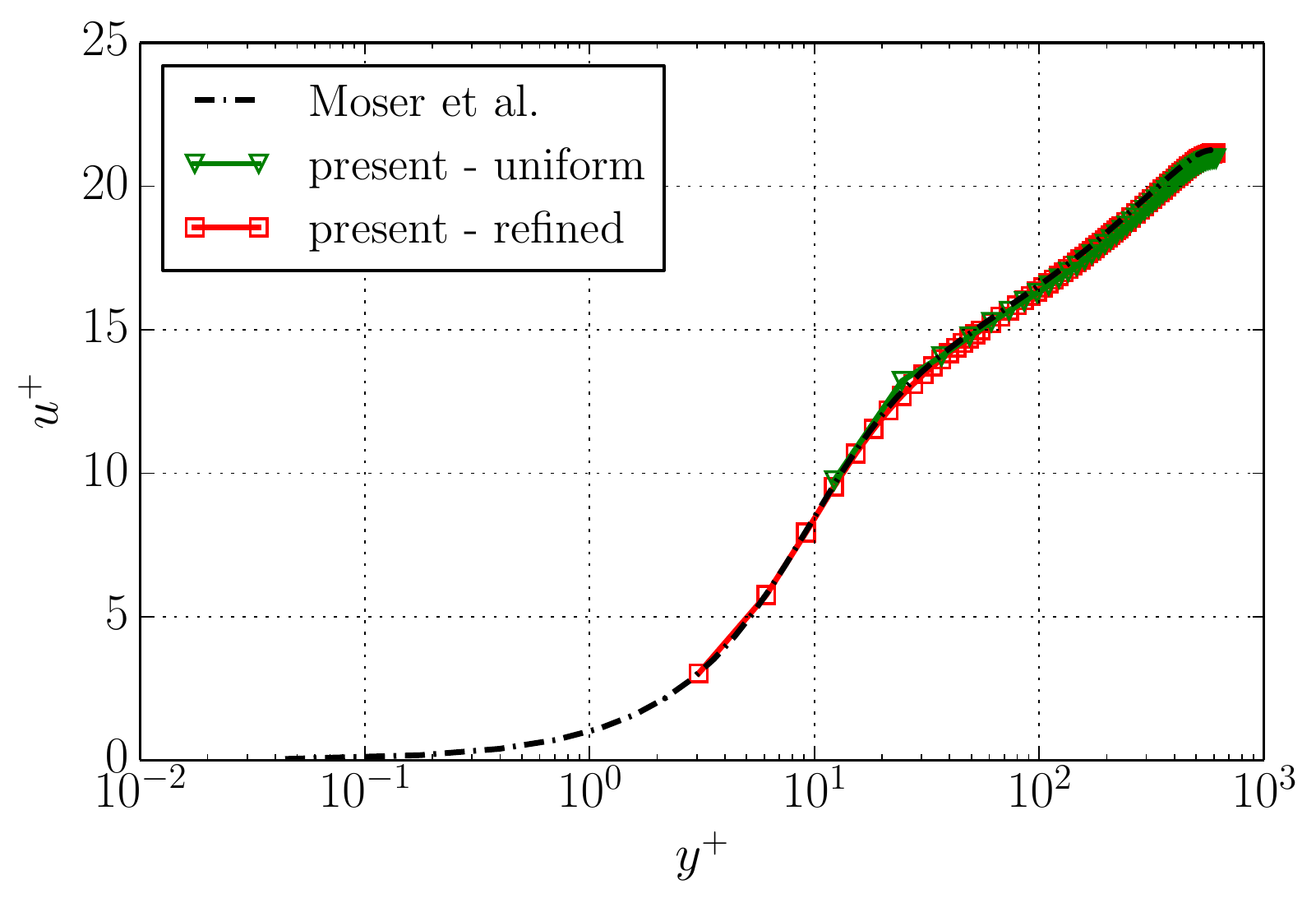}
	\caption{Mean velocity profile in a turbulent channel at $\rm{Re}_\tau=180$}
	\label{fig:channel_u_Re590}
\end{figure}

\begin{figure}
	\centering					
	\includegraphics[width=0.45\textwidth]{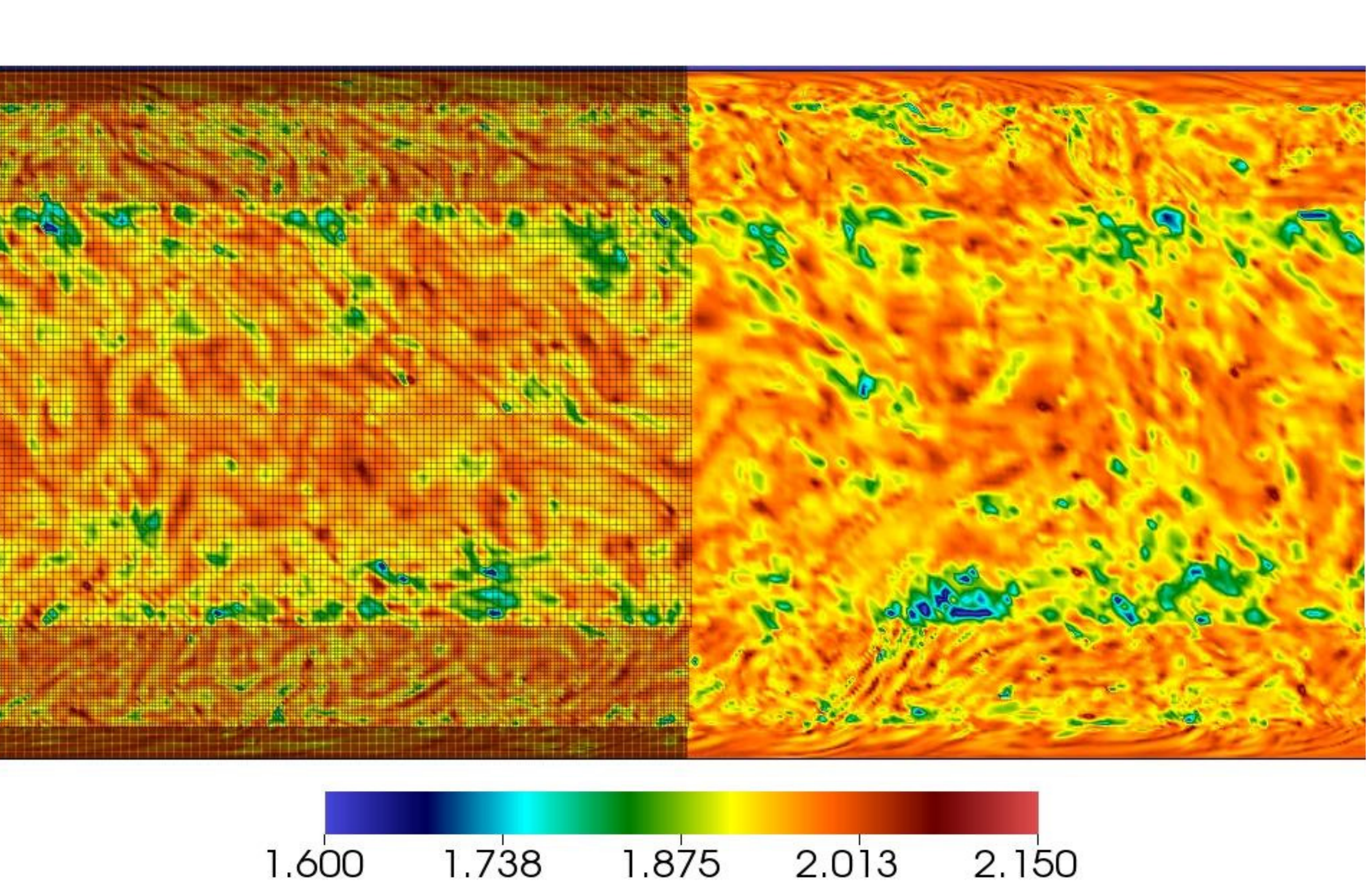}
	\caption{Slice through the turbulent channel flow at $\rm{Re}_\tau=590$ showing the spatial distribution of the stabilizer $\gamma$ for the KBC model and the refinement patches (only left half shown here).}
	\label{fig:channel_gamma}
\end{figure}

The first validation in the isothermal regime is dedicated to the well-studied 
problem of the turbulent flow in a rectangular channel for which many experimental and numerical 
investigations have been conducted (see, e.g., \cite{Eckelmann1974,Kreplin1979,Kim1987, Moser1999}).  
We compare the performance of the proposed grid refinement technique in combination with {KBC} models to the DNS data of \cite{Moser1999} for a Reynolds number $\rm{Re}_\tau=u_\tau \delta /\nu =180$ and $\rm{Re}_\tau=590$.
The friction Reynolds number $\rm{Re}_\tau$ is based on the channel half-width $\delta$ and the friction velocity $u_\tau=\sqrt{\tau_w/\rho}$.
The flow is driven by a constant body force, which was chosen according to $g=\rm{Re}_\tau^2 \nu^2/\delta^3$ to achieve the desired Reynolds number.
By computing the wall-shear stress directly from the flow field of the simulation, the friction velocity may be evaluated 
and the actual Reynolds number measured. The results of our simulations are given in Table \ref{tab:channelData}.
The simulations were conducted for a resolution of the channel half-width of $\delta_c = 50$ lattice points on the coarse level. 


{The computational domain was chosen as [$4\pi\delta \times \delta \times 4/3\pi\delta]$ for $\rm{Re}_\tau=180$ and $[ 8\delta \times \delta \times 4\delta]$
for $\rm{Re}_\tau=590$, where the
$x$- and $z$-coordinate are in streamwise and spanwise direction, respectively.}
An initial perturbation is introduced into the flow field in order to trigger turbulence.
A snapshot of the velocity magnitude for both configurations is shown in Fig.~\ref{fig:channel_velSlice}.
Most conveniently, all data is expressed in wall-units, where the velocity is defined as  $u^+ = u /u_\tau$ and spatial coordinate as $y^+ = y u_\tau/\nu$.
Using wall-units, the spatial resolution may be quantified with the non-dimensional grid spacing $\Delta y ^+$ near the channel wall.
The scaling of the average velocity is well understood in a high-Reynolds number turbulent channel by the law of the wall, where
one distinguishes between the viscous sublayer ($y^+<5$), the buffer layer ($5<y^+<30$) and log-law region ($y^+ \geq 30$) \cite{pope2000turbulent}.
While the average streamwise velocity $u^+$ is assumed to scale linearly with the wall coordinate $y^+$ in the viscous sublayer, 
the log-law suggests a scaling with $u^+ =\kappa^{-1} \ln y^+ + C^+$ in the log-law region, where $\kappa \approx 0.41$ denotes the von K\'arm\'an constant and the constant $C^+$ is given by $C^+ \approx 5.5$. 
While in the $Re_\tau=180$ case low Reynolds number effects may still be observed, 
the channel flow at $\rm{Re}_\tau=590$ is at a sufficiently high Reynolds number
to exhibit all expected features of high Reynolds number wall-bounded flows.

Considering the case of $Re_\tau=180$, we choose one level of refinement near the wall with a spatial extent of $20$ coarse-level points, 
which yields a resolution of {$\Delta y_{f}^+ =1.73$} near the wall. 
We compare the average velocity profile in Fig.~\ref{fig:channel_u_Re180}.
It is evident that the results match the reference data excellently.
Considering the rms velocity profiles, a similarly good agreement is shown in Fig.~\ref{fig:channel_urms_Re180}.
The marginal overshoot of $u_{rms}$ is attributed to a slightly higher Reynolds number in our simulation compared to the reference data.
{Important to notice is that the transition between the grid levels is smooth for both the mean and rms velocity profiles and no numerical artifacts 
are present.}
y

With these results, we test the limits of applicability of the proposed grid refinement technique and choose a higher Reynolds number of $Re_\tau=590$.
In this case, we add an additional grid level in the near-wall region to resolve the flow field in the fine level 
with $\Delta y_f^+ \approx 3.06$ (see Fig.~\ref{fig:channel_gamma}).
To access the necessity of grid refinement in the near-wall region, we conduct another simulation for which a uniform mesh with $\Delta y^+ \approx 12.26$ is used. 
The average velocity profiles are shown in Fig.~\ref{fig:channel_u_Re590}. One can observe that despite of the severe under-resolution 
in the uniform case, the average profile agree well with the reference data. The simulation using the refined mesh matches similarly well
and one cannot observe any discontinuities at the level interfaces. 
Studying this in more detail, we consider the next order of statistics, the rms velocity profiles, for both cases in 
Fig.~\ref{fig:channel_urms_Re590}.
In the uniform case (Fig.~\ref{fig:channel_urms_Re590}, right), it is obvious that the peak $u_{rms}$ is severely under-predicted, yielding a rather poor agreement with the reference DNS. This is of course expected as the small-scale structures are not well represented on such a coarse grid.
Surprisingly, the periodic and streamwise components of the rms velocities show rather small discrepancies when compared to the DNS.
{This is attributed to the excellent subgrid features of entropic lattice Boltzmann models for under-resolved simulations as also reported in \citep{Bosch2015, Dorschner2016}}.

The simulation using grid refinement is shown on the left side of Fig.~\ref{fig:channel_urms_Re590}, where we zoom in the near-wall region and pay special attention to the level interface.
The agreement on the finest level is good, as expected. 
However, a small but distinct jump may be observed at the interface of level one and level two, which
is particularly pronounced for $v_{rms}$. 
This owes to the fact that the severe under-resolution in the coarse level leads to a misrepresentation of the small scale fluctuation 
on the coarse level. To support this hypothesis, in Fig.~\ref{fig:channel_gamma}, we show an instantaneous snapshot of the spatial distribution of the stabilizer $\gamma$ for the KBC model as 
computed by Eq.~(\ref{eq:gamma_min_approx}). 
As shown in \cite{Bosch2015}, the stabilizer $\gamma$ tends to the LBGK value $\gamma=2$ in the resolved case. 
Thus, its deviation provides a measure of under-resolution.
It is apparent that in the refined near-wall region, the stabilizer is indeed very close to $\gamma=2$, 
whereas the distribution in the bulk exhibits large deviations.
It is interesting to notice that particularly large deviations arise in the immediate neighborhood of the level interface. 
It seems that the level interface is detected and compensated by the KBC model, thus rendering explicit projection of the fine level solution 
onto the coarse mesh by filtering or alike unnecessary \citep{Lagrava2012}.\\

These observations when viewed together with the flow at $\rm{Re}_\tau=180$ suggest that despite of the excellent subgrid features of entropic lattice Boltzmann models in matching the average velocity (see uniform mesh at $\rm{Re}_\tau=590$), a minimum resolution on the coarse grid is required to reasonably represent the small scale fluctuations on the coarse grid and thus achieve a smooth grid level transition in higher-order statistics. The implicit subgrid model however, alleviates
the need for filtering the fine level solution on the coarse level and assures stability in the coarse level.
%
{It further needs to be emphasized that due to these excellent subgrid features, the finest patch may be itself under-resolved ($\Delta y_f^+ \approx 3.06$ compared to $\Delta y^+ \approx 0.04$ for the DNS) while accurately capturing the velocity fluctuation near the wall. 
{Thus, the refinement allows to enhance the subgrid features, so that also the higher-order statistics can be captured accurately. }}


\subsubsection{\label{sec:flow_past_sphere}Flow past a sphere}
\begin{figure}[!t]
	\centering					
	\includegraphics[width=0.49\textwidth]{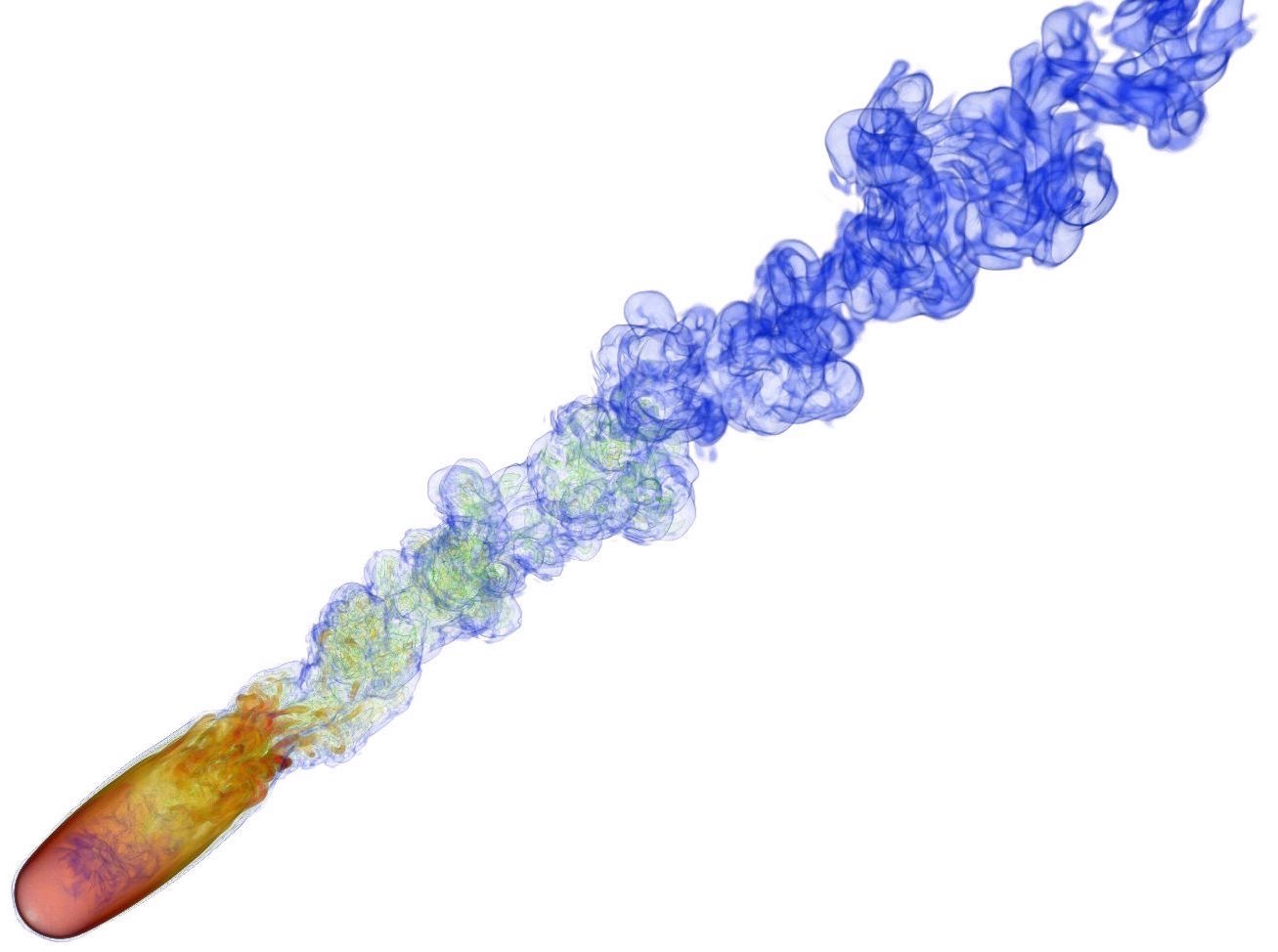}
	\caption{Vorticity volume rendering for the flow past a sphere at $\rm{Re}=3700$.}
	\label{fig:volumeRendering_sphere}
\end{figure}
\begin{figure*}
	\centering					
	\includegraphics[width= 0.9\textwidth]{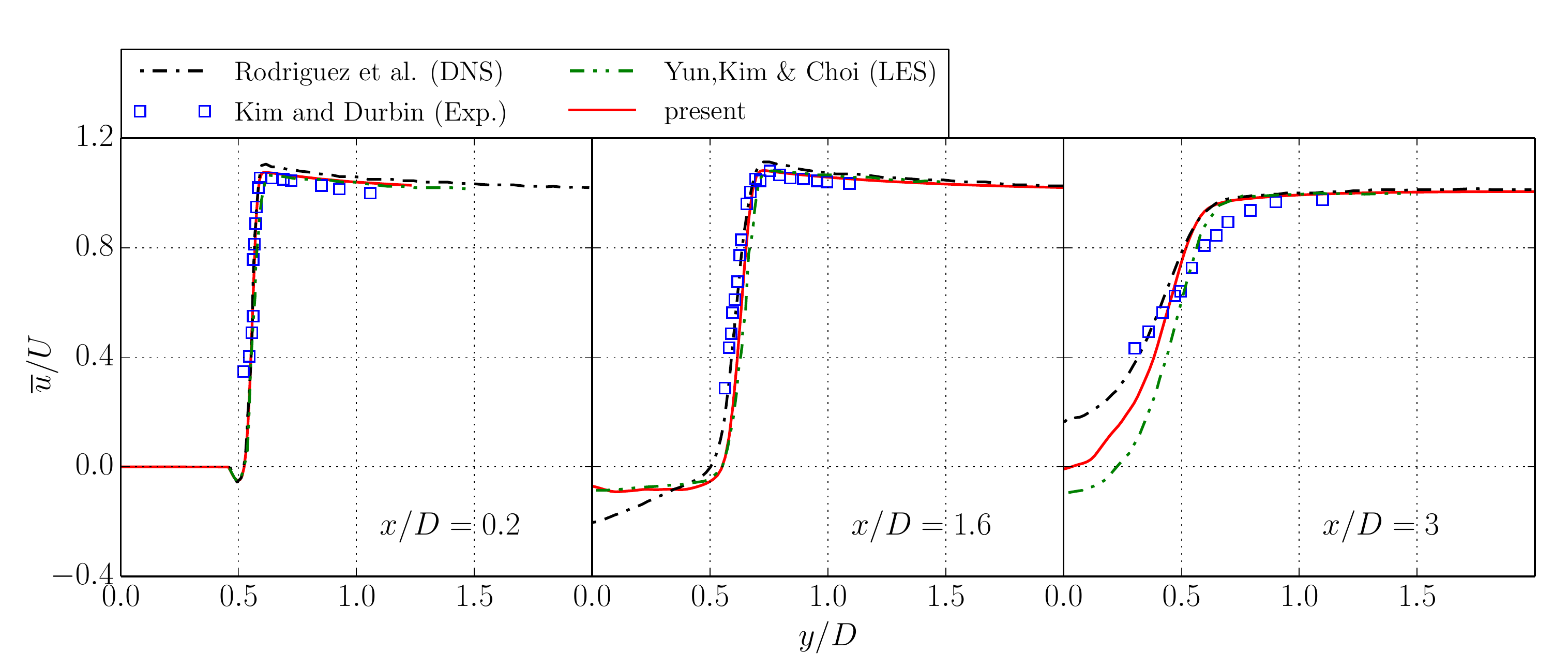}
	\caption{Mean streamwise velocity profiles in the wake for the simulation of flow past a sphere at $\rm{Re}=3700$. }
	\label{fig:sphere_R3700_velProfiles}
\end{figure*}

\begin{figure}
	\centering					
	\includegraphics[width=0.49\textwidth]{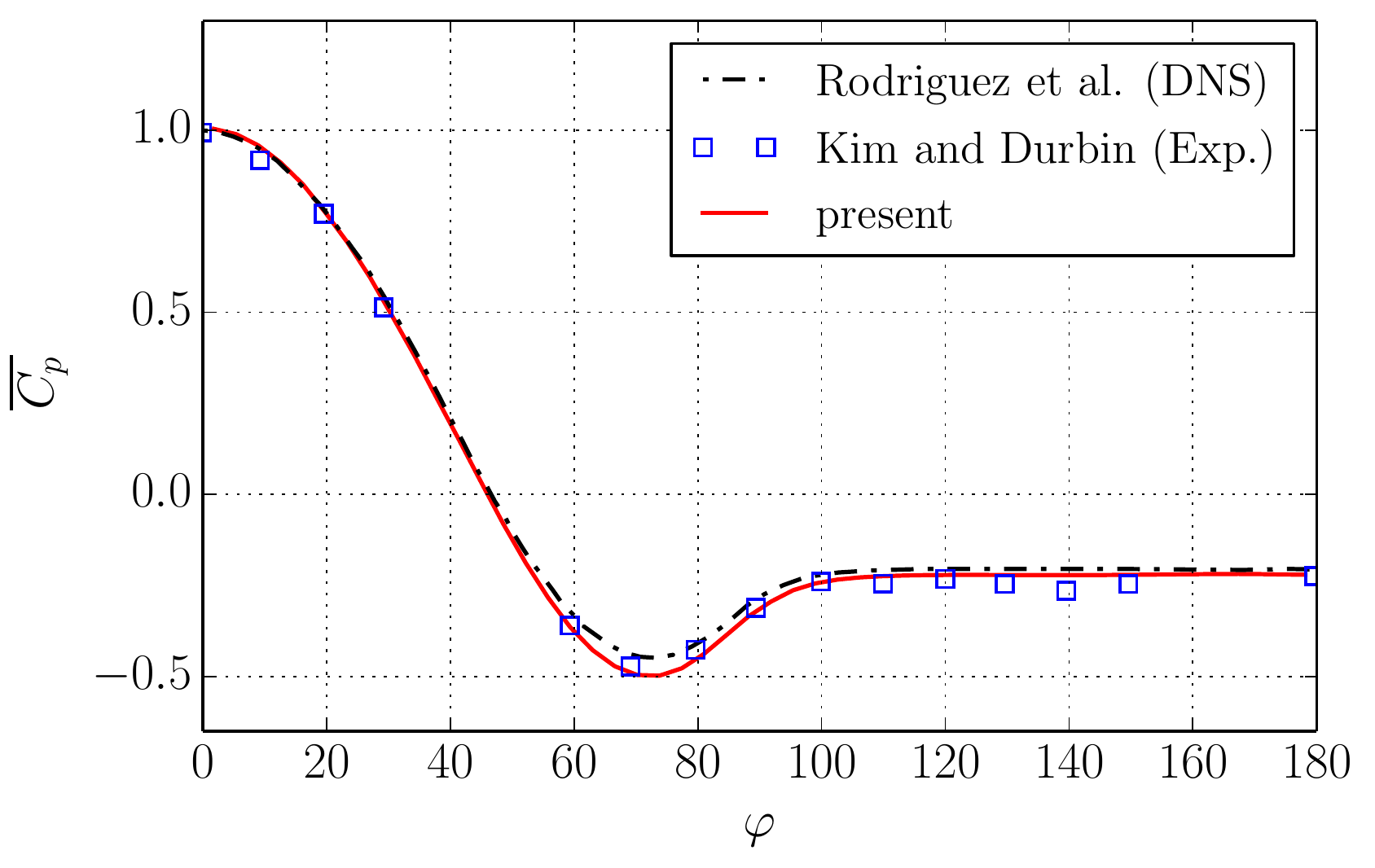}
	\caption{Pressure coefficient distribution around the sphere at $\rm{Re}=3700$. }
	\label{fig:sphere_R3700_cp}
\end{figure}

After the validation of the grid refinement technique for flat walls in the previous section, we now consider curved walls and
choose the well-studied problem of the flow past a sphere. 
In order to access the accuracy of the proposed grid refinement algorithm, we
focus on the flow within the sub-critical regime, for which the boundary layer remains 
laminar and the near wake features turbulence. 
A detailed analysis is conducted for $\rm{Re}=3700$ and an instantaneous snapshot of the vorticity volume rendering is 
shown in Fig.~\ref{fig:schematic}.
The computational domain is chosen as $[-7D,23D] \times [-10D, 10D] \times  [-10D, 10D]$ with 
the sphere centered at the origin.
Four refinement patches are located closely around the sphere, where the finest patch resolves the sphere with $D=120$.
It is worth noting that a simulation for this sphere resolution and without grid refinement would require $N_f \approx 20.7 \cdot 10^9$ grid points, rendering such
a simulation unfeasible for practical purposes.
{With the refinement, the computational grid reduces to a total of $ N_r \approx 94.6 \cdot 10^6$ lattice points.}
By taking into account the time step scaling, the equivalent fine level points in the refined case amount to 
$39.5 \cdot 10^6$, which is roughly $575$ times less than the fully resolved case without refinement.
Such an estimate suggest a tremendous optimization potential when employing grid refinement, while still retaining the desired accuracy. This
allows for a detailed comparison with the contributions of \cite{Rodriguez2011} and their DNS simulation, \cite{Yun2006} performing a LES simulation
and the experimental results of \cite{KimH.J.Durbin1988,schlichting2003}.
First, we compare various scalar quantities such as the mean drag coefficient $\overline{C_d}=\overline{F_d} /(1/2\rho_{\infty} U^2 D)$, the mean base pressure coefficient 
$\overline{C_{pb}}$, the recirculation length $\overline{L_r}$ and the separation angle $\overline{\varphi_s}$. Here, the drag force is denoted by $F_d$.
As tabulated in Table \ref{tab:sphere_Re3700}, the comparison shows excellent agreement for all quantities with the literature. 
\begin{table}[!b]
\caption{\label{tab:table4}%
\label{tab:sphere_Re3700}
Turbulent flow past sphere at $\rm{Re}=3700$ and the comparison of the mean drag coefficient $\overline{C_d}$, the averaged base pressure coefficient $\overline{C_{pb}}$,
the recirculation length $\overline{L_r}$ and the separation angle $\overline{\varphi_s}$ with literature values.}
\begin{ruledtabular}
\begin{tabular}{lcccc}
Contribution						&	$\overline{C_d}$		&	$\overline{C_{pb}}$		&	$\overline{L}_r$			&	$\overline{\varphi_s}$ 	\\
\hline
\citet{Rodriguez2011} 			&  $0.394$				& 	$-0.207$					& $2.28$						&	$89.4$		\\	
\citet{Yun2006} 					&  $0.355$				& 	$-0.194$					& $2.622$						&	$90$		\\	
\citet{KimH.J.Durbin1988} 	&  $-$						& 	$-0.224$					& $-$							&	$-$			\\	
\citet{schlichting2003}		&  $0.39$					& 	$-$							& $-$							&	$-$			\\	
present							&  $0.383$				&	$-0.220$					& $2.51$  					&  $89.993$
\end{tabular}
\end{ruledtabular}
\end{table}



For a more thorough analysis, we compare the time-averaged profiles of the streamwise velocity component $\overline{u}$ in the near wake to profiles obtained by DNS, LES and experiment, see Fig.~\ref{fig:sphere_R3700_velProfiles}.
Three profiles are measured for a streamwise location of $x/D=0.2$, $x/D=1.6$ and further downstream at $x/D=3$.
While for $x/D=0.2$ all measurements are in almost perfect agreement, the discrepancies increase slightly for all reference data further downstream.
Nonetheless, the measurements taken from our simulation appear to be in good agreement with all reference data.
Next, in Fig.~\ref{fig:sphere_R3700_cp}, the azimuthally averaged distribution of the mean pressure coefficient 
$\overline{C_p}=(\overline{p} - p_{\infty})/(1/2 \rho_\infty U^2)$ around the sphere is presented in comparison with the DNS and experimental results.
It is apparent that mean pressure distribution matches the two references well.
However, it needs to be pointed out that this agreement could only be achieved with an additional layer of refinement, yielding a resolution 
of $D=240$ points for the diameter of the sphere.
This is in contrast to the velocity profiles, which were captured already with four levels and a diameter of $D=120$ points in the finest level.

These results for the turbulent channel flow and the flow past a sphere conclude the validation of the isothermal regime. Our results indicate robustness, high accuracy and compatibility with entropy-based LBM of the proposed grid refinement scheme for high Reynolds number turbulent flows. 
{Although average flow velocity is easily captured using a coarse uniform grid in combination with the KBC model, near-wall features and higher-order statistics do require grid refinement for an accurate representation.}

\subsection{Thermal flows}

We proceed investigating stability and accuracy of the proposed grid refinement algorithm when applied to thermal flows, simulated with the two-population model \cite{Karlin2013}. We start with the simulation of Rayleigh-B\'enard convection (RBC) in order to validate the model and the grid refinement algorithm for flat walls.
{As a second step, in Sec.~\ref{sec:flow_past_sphere}, we revisit the simulation of the flow past a sphere but additionally include the temperature field, and compare the mean Nusselt number distribution.}
The boundary conditions used for all wall-boundary nodes is Grad's approximation as presented in \cite{pareschi2016thermal}.

\subsubsection{\label{sec:rayleigh_benard_convection }Rayleigh-B\'enard convection}

\begin{figure}
	\centering					
	\includegraphics[trim=2.8cm 3cm 2.2cm 3cm, clip, width=0.5\textwidth]{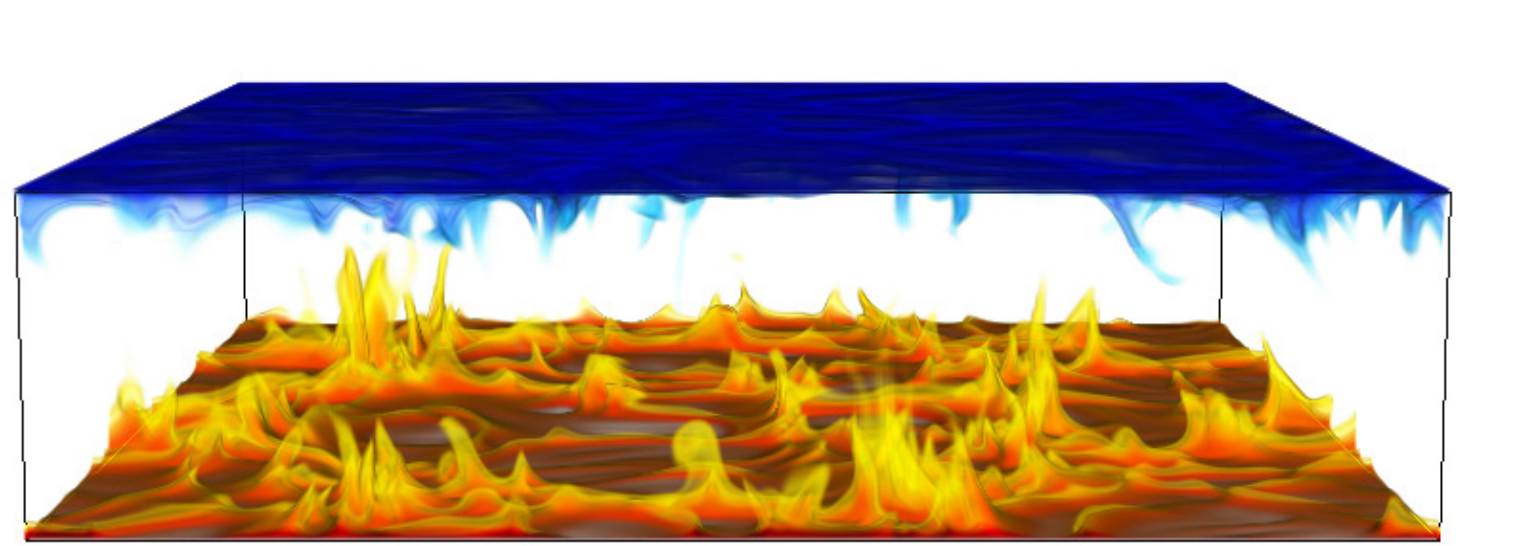}
	\caption{Volume rendering of {temperature for} the Rayleigh-B\'enard convection at $\rm Ra = 10^7$.}
	\label{fig:RBC_volume_render}
\end{figure}

\begin{figure}
	\centering					
	\includegraphics[width=0.5\textwidth]{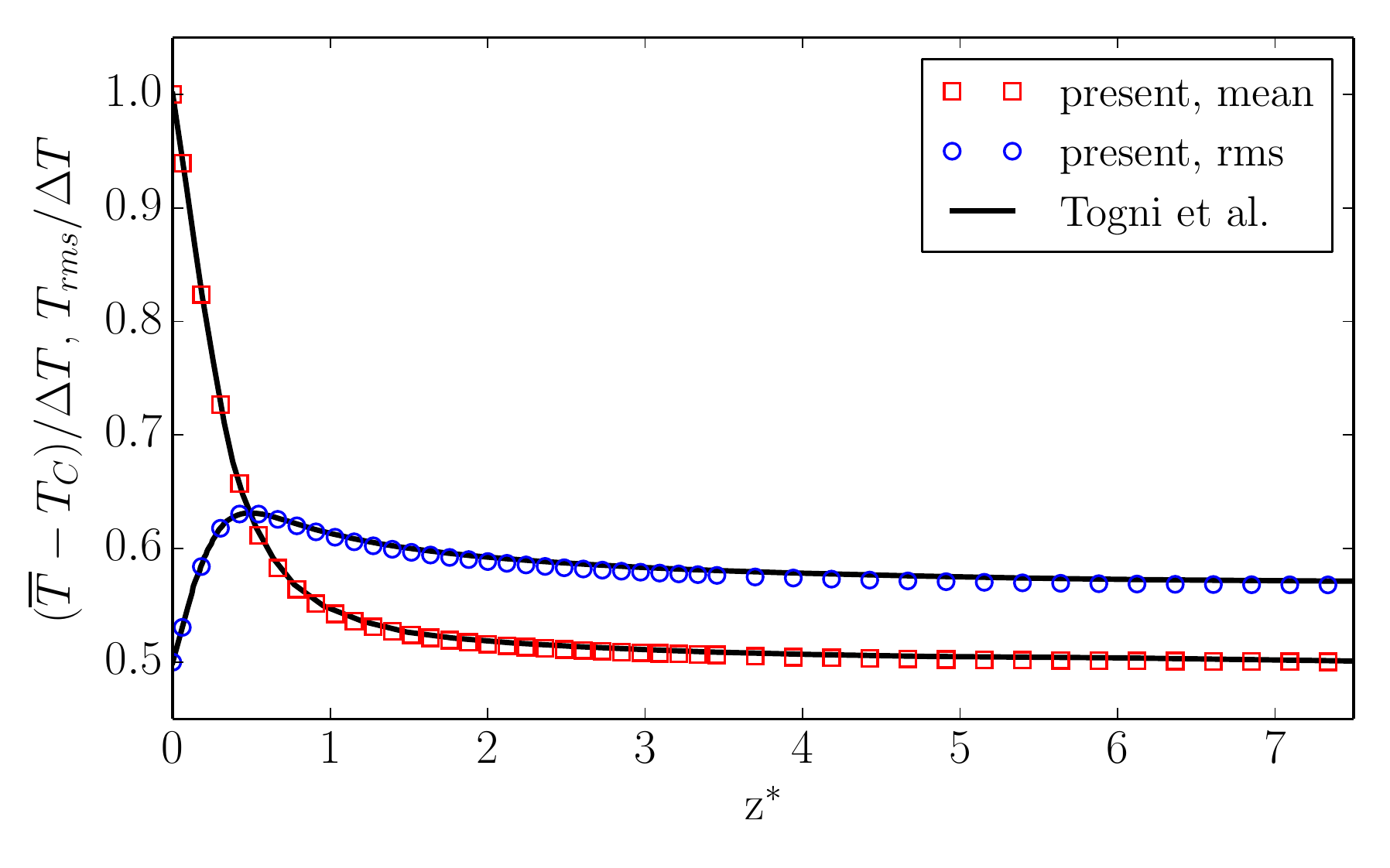}
	\caption{Comparison of mean and rms temperature profiles for the Rayleigh-B\'enard convection at $\rm Ra = 10^7$. The rms temperature profile is shifted by $0.5$ in order to ease visualization of the plot.}
	\label{fig:RBC_T_Ra1e7}
\end{figure}

\begin{figure}
	\centering					
	\includegraphics[width=0.5\textwidth]{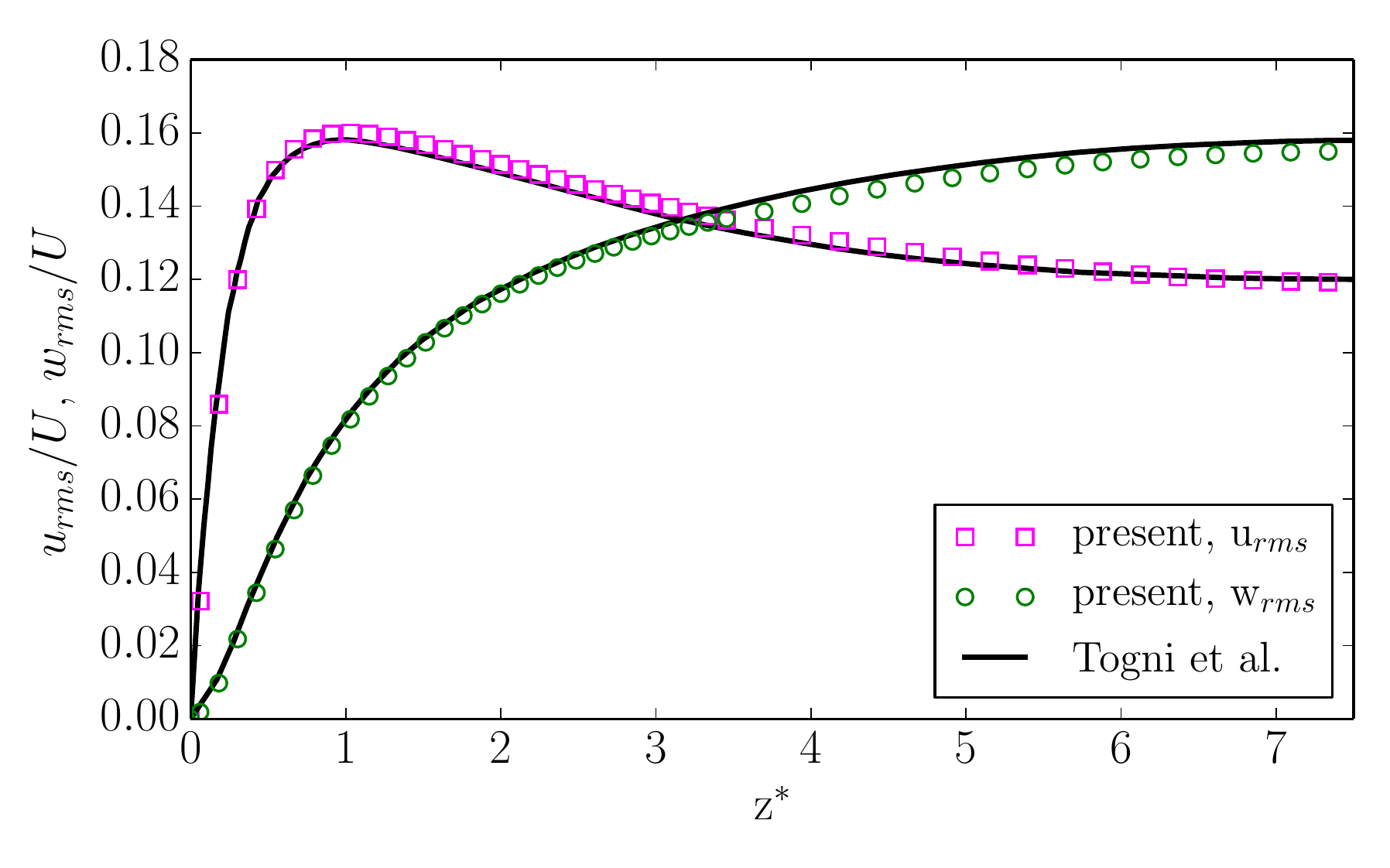}
	 \caption{Comparison of rms u- and w-velocity components profiles for the Rayleigh-B\'enard convection at $\rm Ra =10^7$.}
	\label{fig:RBC_u_w_rms_Ra1e7}
\end{figure}

The Rayleigh-B\'enard set-up consists of a fluid layer which is heated from below and cooled from above. When the temperature difference $\Delta T$ between the two walls is sufficiently high, thermal convection is triggered. The non-dimensional parameters governing this problem are the Rayleigh number $\rm Ra = g \lambda \Delta T H^3/\nu \kappa $ and the Prandtl number $\rm Pr = \nu/\kappa $, where $g$ represents gravitational acceleration, $\lambda$ the thermal expansion coefficient, $H$ the height of the fluid layer, $\nu$ the kinematic viscosity and $\kappa$ the thermal conductivity. \\
In this work, we present the results of the thermal two-population model coupled with the grid refinement algorithm for thermal flows and their comparison with the DNS simulation of \cite{togni2015physical}. The Rayleigh and Prandtl number are $\rm Ra = 1\cdot 10^7 $ and $\rm Pr = 0.7$. 
{The computational domain is a box with periodicity in the x- and y-directions, where a fixed temperature is imposed at the bottom (hot) and top (cold) walls.} The size of the domain is $L_x \times L_y \times L_z = 8H \times 8H \times H$, where $H = 64$ is the resolution in the coarse grid level. On top of the coarse grid, one more refinement patch is added at both hot and cold walls in order to increase the resolution in the boundary layers. 
To trigger convection an initial random perturbation is imposed on a linear temperature profile. {The buoyancy force is computed according to the Boussinesq assumption, and it is implemented as reported in \cite{Karlin2013}}. \\
An instantaneous volume rendering of the temperature in the domain is shown in Fig.~\ref{fig:RBC_volume_render}.
One can notice the cold temperature plumes in the upper part of the box and the hot temperature plumes developing from the bottom of the box.
Quantitatively, in Fig.~\ref{fig:RBC_T_Ra1e7}, we compare the mean and the rms temperature profiles with the recent DNS data 
\cite{togni2015physical} and the agreement is good.
{All statistics are collected after the initial transient at $40t_L$ and sampled every $2.5 t_L$ for a 
time period of $100 t_L$, where $t_L=2H/U$ denotes the large eddy turnover time with the free fall velocity $U = \sqrt{g \lambda \Delta T H}$.
The temperature profiles are plotted as a function of the normalized vertical coordinate (normal to the bottom wall) $z^{*} = z/(H/\rm{\overline{Nu}})$, where $\rm{\overline{Nu}}$ is the mean Nusselt number, defined as $\rm{ \overline{Nu}}=\frac{d \overline{T} / d n }{\Delta T / H}$, where $d \overline{T} / d n$ is the mean temperature gradient at top and bottom walls.}
In Fig.~\ref{fig:RBC_u_w_rms_Ra1e7}, we study the rms velocity profiles for the $u$ and $w$ component and compare to the reference data.
Analogous to the temperature profiles, a good agreement is also observed for the velocity fluctuations.
To complete the validation for the Rayleigh-B\'enard convection we compare the resulting Nusselt number. 
While in our simulation the Nusselt number at the wall is evaluated as $\rm{Nu}_{ELBM} = 15.67$, the DNS \cite{togni2015physical} reports $\rm{Nu}_{DNS} = 15.59$.
This amounts to a relative discrepancy of $\epsilon_{\rm Nu} = 5.13 \cdot 10^{-3}$. \\

%
%
%
\begin{figure}
	\centering					
	\includegraphics[trim=4.5cm 2.0cm 1.5cm 3cm, clip, width=0.5\textwidth]{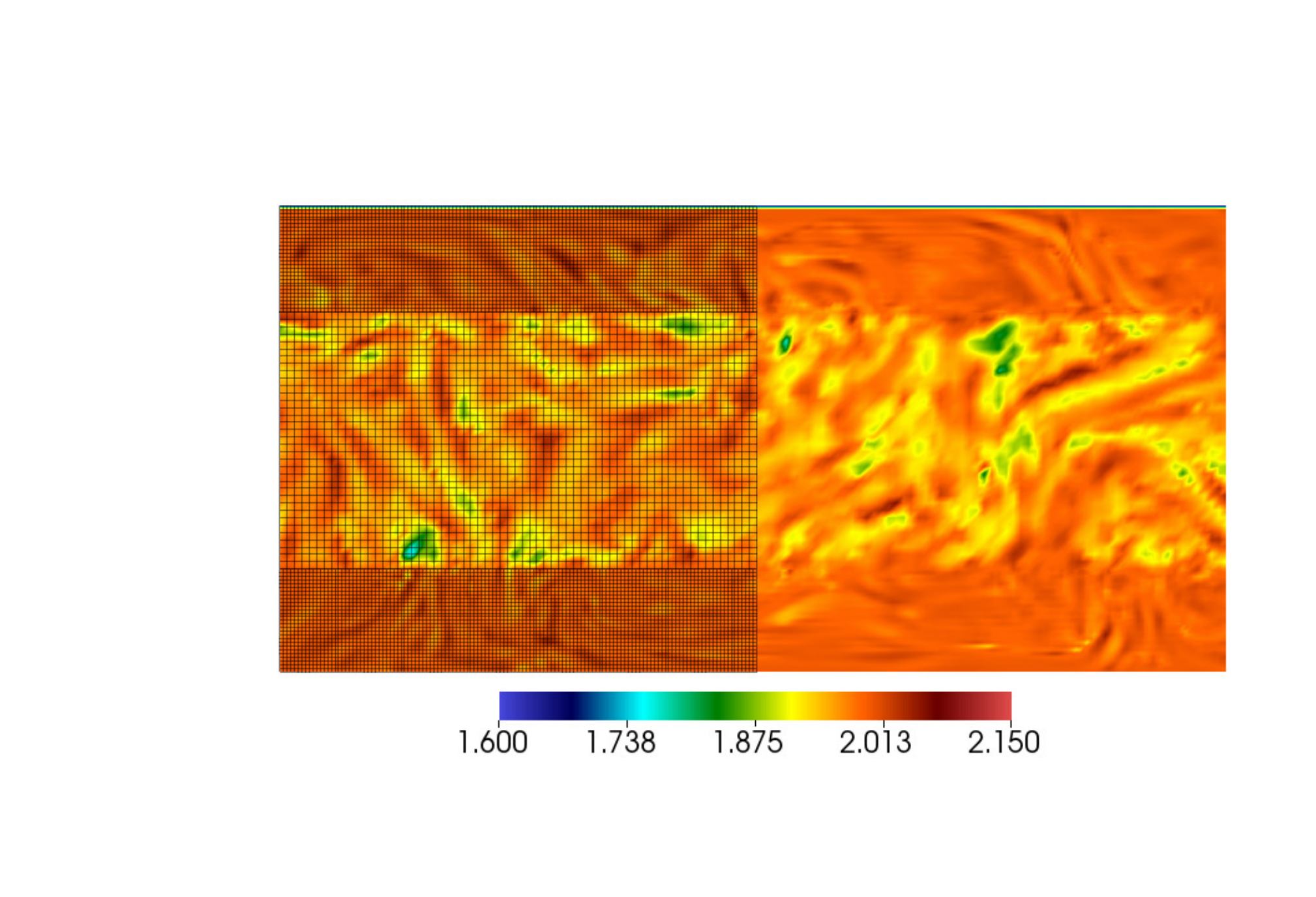}
	\caption{Slice through the Rayleigh-B\'enard convection at $\rm Ra = 1\cdot 10^7$ showing the spatial distribution of the stabilizer $\gamma$ for the KBC model and the refinement patch (only left half shown here).}
	\label{fig:RBC_gamma}
\end{figure}

{Similar to the turbulent channel flow, it is important to notice that the transition at the grid interface is smooth for all quantities} presented here, from the mean and rms temperature profiles to the rms x- and z-velocity profiles. This is again attributed to the entropic stabilizer $\gamma$ for which a snapshot through the domain is shown in Fig.~\ref{fig:RBC_gamma}. It is evident that larger deviations of $\gamma$ from the LBGK value $\gamma = 2$ appear in the bulk rather than near the walls where the resolution is higher. Also in this case larger deviations in the $\gamma$ arise in the immediate neighborhood of the level interface, which seems to be detected and compensated by the KBC model. This is analogous to the observation in the channel flow.  
{Thus demonstrating the self-adaptive nature of the entropic stabilizer $\gamma$ for all flow situations including grid refinement and complex wall boundary conditions. This self-adaptive nature of $\gamma$ makes the simulations parameter-free.}

\subsubsection{Flow past a heated sphere}

\begin{figure}
	\centering					
	\includegraphics[trim=4cm 0cm 1.5cm 2.5cm, clip, width=0.5\textwidth]{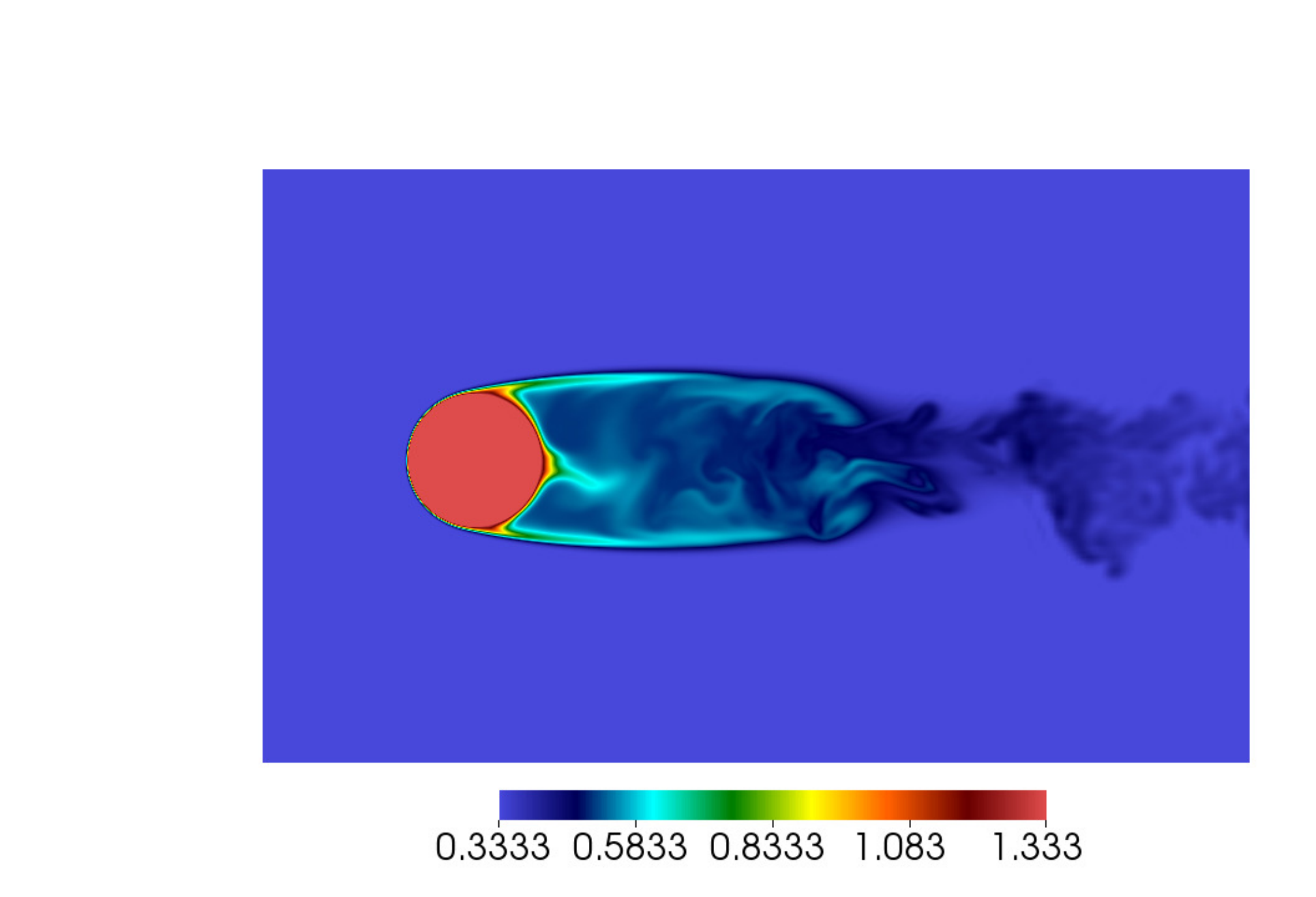}
	\caption{Snapshot of the instantaneous  temperature distribution around and in the wake a heated sphere at $\rm Re=3700$.}
	\label{fig:sphere_T_distribution}
\end{figure}

\begin{figure}
	\centering					
	\includegraphics[width=0.49\textwidth]{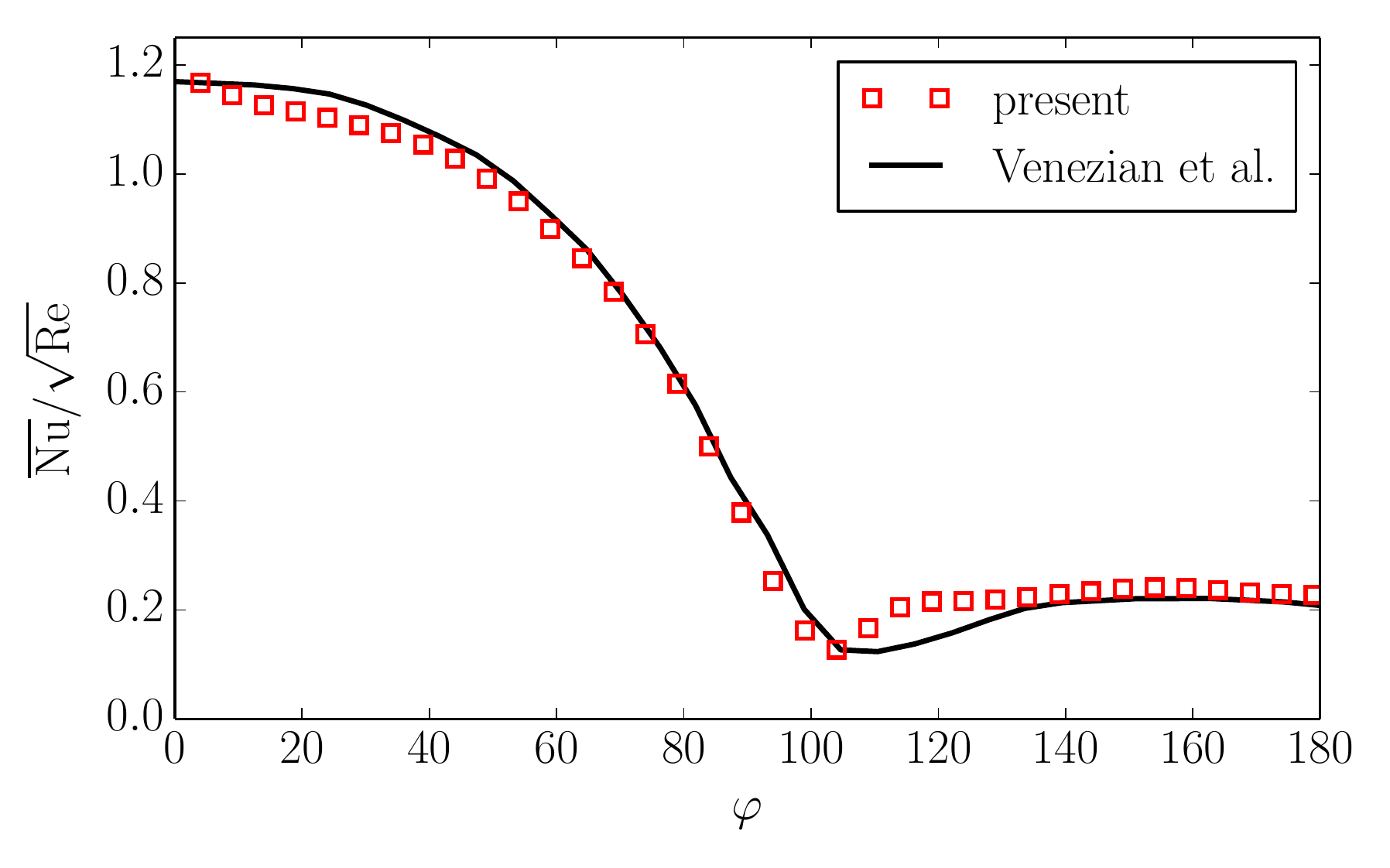}
	\caption{Average Nusselt number distribution around the heated sphere at $\rm{Re} = 3700$.}
	\label{fig:sphere_Nu_Re3700}
\end{figure}

As in the isothermal section, after validation of the scheme for flows involving flat walls, we increase the complexity and consider
curved walls next.
For this purpose, we consider a heated sphere with the surface at constant temperature. The flow is simulated for a Reynolds number $\rm{Re} = 3700$ and Prandtl number $\rm Pr = 0.7$. 
The computational domain and the refinement patches are identical to the isothermal case with $D=240$.
A snapshot of the temperature distribution around and in the wake of the sphere for this simulation is shown in Fig.~\ref{fig:sphere_T_distribution}.
{The figure shows the hot temperature streams in the shear layer and the back of the sphere, where the flow recirculates.} Further downstream, hot fluid mixes with cold fluid and the temperature becomes diluted. In order to quantitatively validate the heated sphere simulation, the mean reduced Nusselt number distribution $\overline{\rm{Nu}}/\sqrt{\rm Re} = \frac{d \overline{T} / d n }{\Delta T / D}/\sqrt{\rm Re}$ around the sphere is shown in comparison to experimental results \cite{venezian1962thermal} in Fig.~\ref{fig:sphere_Nu_Re3700}, where $n$ is the normal coordinate with respect to the sphere surface and $\Delta T$ the temperature difference between fluid and sphere surface.
The plot shows a good comparison with the experiment.

\subsection{Compressible flows}

We conclude the numerical validation by entering the compressible regime for which we take the simulation of the two-dimensional viscous supersonic flow around a NACA0012 airfoil as an example.

\subsubsection{Supersonic NACA0012 airfoil}

\begin{figure}
	\centering					
	\includegraphics[trim=3.5cm 0cm 3cm 1.4cm, clip, width=0.48\textwidth]{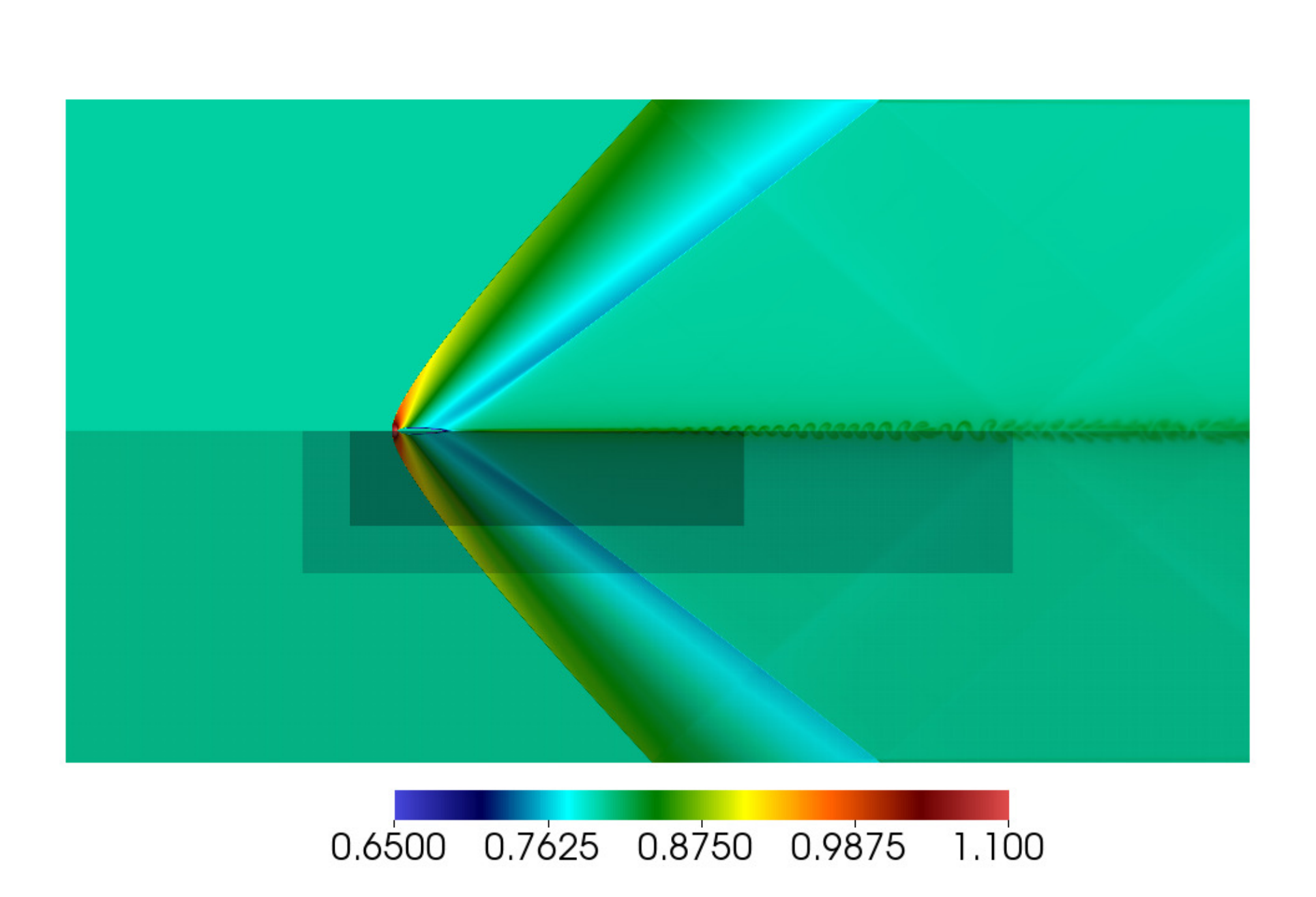}
	\caption{Snapshot of temperature distribution around the NACA0012 airfoil at $\mathcal{A} = 0^{{\circ}}$, $\rm Ma = 1.5$ and $\rm Re = 10000$.  Grid refinement patches are shown by the shadowed regions around the airfoil in the lower half of the domain.}
	\label{fig:NACA0012_snapshot_T}
\end{figure}

The set-up consists of a two-dimensional simulation of the viscous supersonic flow field around a NACA0012 airfoil, at zero angle of attack $\mathcal{A} = 0^{\circ}$. The free-stream Mach number is set to $\rm{Ma_{\infty}} = U_{\infty}/a_{\infty} = 1.5$, where $a_{\infty} = \sqrt{\gamma_{ad} T_{\infty}}$ is the speed of sound with $\gamma_{ad} = 1.4$ and $T_{\infty} = 0.8$,  while the Reynolds number, based on the chord $C$ of the airfoil, is $ \rm{Re} = CU_{\infty}/\nu= 10000$.
The computational domain is {prescribed as} $[-7.5C,17.5C] \times [-7C, 7C]$ with the airfoil centered at the origin. Two refinement patches are placed closely around the airfoil, where the finest patch resolves the airfoil chord with $C=1200$ grid points (see Fig.~\ref{fig:NACA0012_snapshot_T}). \\
{Due to the high Mach number in this simulation, we use the shifted lattices as presented in \cite{frapolli2016lattice}. In particular, we employ the D$2$Q$49$ lattice with a shift in the freestream direction of $U_x=1$. 
The advantage of shifted lattices is that the errors in the higher-order moments of the equilibrium populations are also shifted and centered around the shift velocity $U_x$.
This allows us to keep the number of populations of the multispeed lattice relatively small while reducing the errors in the high Mach regime. For further details on shifted lattices the reader is referred to \cite{frapolli2016lattice}.} 

In Fig.~\ref{fig:NACA0012_snapshot_T}, a snapshot of the temperature distribution along with the two refinement patches around the airfoil, 
indicated by the shadowed regions in the lower half of the domain, is shown. The main features of the viscous supersonic flow field are evident from the temperature distribution: 
In front of the airfoil, the formation of a bow shock may be observed, yielding a jump and drastic increase in temperature. At the trailing edge of the airfoil, an oblique shock wave develops from the shear layer as a lambda shock. Further downstream, in the shear layer, vortex shedding is initiated. 
Important to notice is that the shock waves penetrate through various refinement levels, 
where special care usually needs to be taken to avoid artificial reflections at the interface.
It is apparent that for the proposed refinement algorithm, no reflections or similar artifacts are observed and thus making it 
suitable also for compressible flows and the related shock dynamics.
%
%

\begin{figure}
	\centering					
	\includegraphics[width=0.50\textwidth]{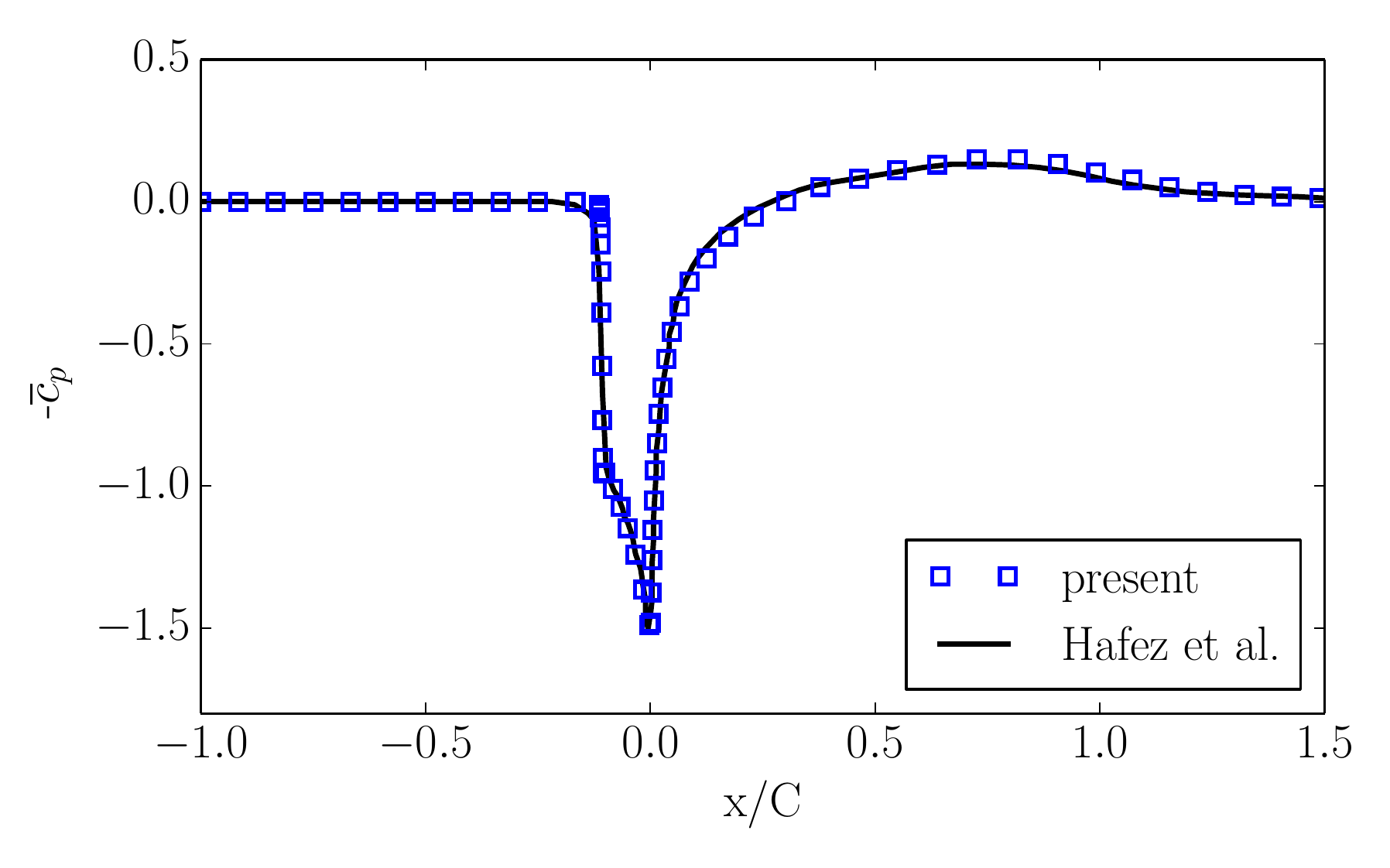}
	\caption{Comparison of the pressure coefficient distribution upstream the airfoil ($x/C \in \left[-1,0 \right]$), on the airfoil surface ($x/C \in \left[0,1 \right]$) and downstream the airfoil ($x/C \in \left[1,1.5 \right]$) for the supersonic simulation of the flow around a NACA0012 airfoil at $\mathcal{A} = 0^{{\circ}}$, $\rm Ma = 1.5$ and $\rm Re = 10000$.}
	\label{fig:cp_NACA0012_Ma15_Re10k_refined_C1200}
\end{figure}

For a quantitative comparison, the mean pressure coefficient distribution upstream the airfoil ($x/C \in \left[-1,0 \right]$), on the airfoil surface ($x/C \in \left[0,1 \right]$) and downstream the airfoil ($x/C \in \left[1,1.5 \right]$), is plotted in Fig.~\ref{fig:cp_NACA0012_Ma15_Re10k_refined_C1200} along with the simulation results reported in \cite{hafez2007simulations}. Statistics have been collected after $50 t^{\ast} = 50 C/U_{\infty}$ flow times, and at every time step in the coarse level.
Evidently, an excellent comparison can be reported. \\

\begin{figure}
	\centering					
	\includegraphics[trim=5cm 1.5cm 4cm 3cm, clip, width=0.50\textwidth]{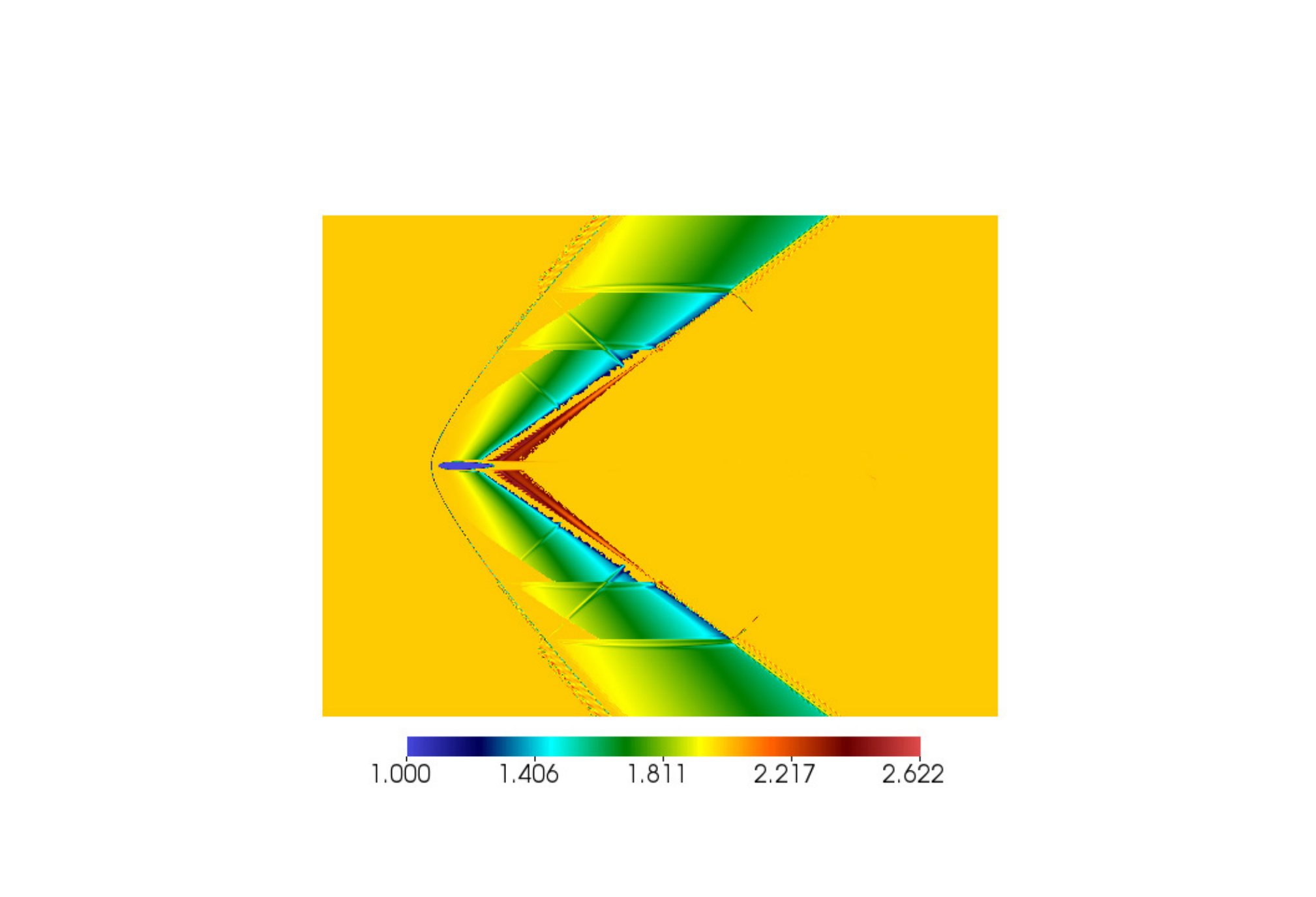}
	\caption{Snapshot of the distribution of the entropic estimate $\alpha$ around the NACA0012 airfoil.}
	\label{fig:NACA0012_Ma15_alpha_distribution}
\end{figure}

Before concluding the numerical validation, we present a snapshot of the entropic estimate distribution 
around the airfoil in Fig.~\ref{fig:NACA0012_Ma15_alpha_distribution}. From the picture, two main observations can be made. 
{First, it is apparent that the entropic estimate adapts to the main physical features of the flow.}
In particular, large deviations from the {resolved limit} value $\alpha = 2$ may be observed near the bow shock, through the expansion wave preceding the oblique shock at the trailing edge, and in the oblique shock itself. A second key observation concerns the interplay of the entropic estimate with the physical feature of the flow and the grid refinement patches. For example, it can be seen that the entropic estimate exhibits larger deviations from the fine to the coarse grid when a shock wave crosses the interface (see interface position in Fig.~\ref{fig:NACA0012_snapshot_T}). 
These deviations in the entropic estimate arises naturally from the entropy condition Eq. (\ref{eq:entropy_condition}) and plays a central role in sustaining the flow field, damping the spurious oscillations near the shock regions or damping the spurious noise, which would would otherwise appear near the different grid interfaces.

\section{Conclusion}
In this paper we have presented a novel grid refinement technique, 
which avoids the low-order time interpolation commonly used in lattice Boltzmann simulations. 
{An extension to thermal and compressible models is achieved by an appropriate rescaling of the populations and thus widening the range of applicability of the proposed grid refinement algorithm. 
Accuracy and robustness is established through various set-ups in the incompressible, thermal and compressible flow regimes for which 
local grid refinement is crucial in order to obtain accurate results at a reasonable computational cost. 
{The implicit subgrid features of entropy-based lattice Boltzmann models render the solutions stable and allow for significant under-resolution while retaining
accuracy.
The entropic stabilizer adapts to the flow features and refinement patches, which enables multi-scale simulations where 
the fine-to-coarse level projection and vice versa is implicit to the model.
%
These features are particularly important in the simulation of the supersonic airfoil, where the shock waves cross the refinement patches without being reflected and destabilizing the flow. 
%
With these insights, an extension of entropic LBMs to adaptive grid refinement seems natural, where
the deviation of the stabilizer from its resolved value is a measure of under-resolution and can thus serve as a refinement criterion. 
This is left for future investigations. 
}
In conclusion, it has been shown that the proposed grid refinement technique in combination with entropy-based lattice Boltzmann models
enables accurate and efficient simulations of flows ranging from low Mach number turbulence all the way to supersonic compressible flows.


\begin{acknowledgments}
This work was supported by the European Research Council (ERC) Advanced Grant
No. 291094-ELBM and the ETH-32-14-2 grant. The computational resources at the Swiss
National Super Computing Center CSCS were provided under the grant s492 and s630.
\end{acknowledgments}



\bibliography{refs}

\begin{thebibliography}{41}%
\makeatletter
\providecommand \@ifxundefined [1]{%
 \@ifx{#1\undefined}
}%
\providecommand \@ifnum [1]{%
 \ifnum #1\expandafter \@firstoftwo
 \else \expandafter \@secondoftwo
 \fi
}%
\providecommand \@ifx [1]{%
 \ifx #1\expandafter \@firstoftwo
 \else \expandafter \@secondoftwo
 \fi
}%
\providecommand \natexlab [1]{#1}%
\providecommand \enquote  [1]{``#1''}%
\providecommand \bibnamefont  [1]{#1}%
\providecommand \bibfnamefont [1]{#1}%
\providecommand \citenamefont [1]{#1}%
\providecommand \href@noop [0]{\@secondoftwo}%
\providecommand \href [0]{\begingroup \@sanitize@url \@href}%
\providecommand \@href[1]{\@@startlink{#1}\@@href}%
\providecommand \@@href[1]{\endgroup#1\@@endlink}%
\providecommand \@sanitize@url [0]{\catcode `\\12\catcode `\$12\catcode
  `\&12\catcode `\#12\catcode `\^12\catcode `\_12\catcode `\%12\relax}%
\providecommand \@@startlink[1]{}%
\providecommand \@@endlink[0]{}%
\providecommand \url  [0]{\begingroup\@sanitize@url \@url }%
\providecommand \@url [1]{\endgroup\@href {#1}{\urlprefix }}%
\providecommand \urlprefix  [0]{URL }%
\providecommand \Eprint [0]{\href }%
\providecommand \doibase [0]{http://dx.doi.org/}%
\providecommand \selectlanguage [0]{\@gobble}%
\providecommand \bibinfo  [0]{\@secondoftwo}%
\providecommand \bibfield  [0]{\@secondoftwo}%
\providecommand \translation [1]{[#1]}%
\providecommand \BibitemOpen [0]{}%
\providecommand \bibitemStop [0]{}%
\providecommand \bibitemNoStop [0]{.\EOS\space}%
\providecommand \EOS [0]{\spacefactor3000\relax}%
\providecommand \BibitemShut  [1]{\csname bibitem#1\endcsname}%
\let\auto@bib@innerbib\@empty
\bibitem [{\citenamefont {Succi}(2015)}]{succi2015lattice}%
  \BibitemOpen
  \bibfield  {author} {\bibinfo {author} {\bibfnamefont {S.}~\bibnamefont
  {Succi}},\ }\href@noop {} {\bibfield  {journal} {\bibinfo  {journal} {EPL
  (Europhysics Letters)}\ }\textbf {\bibinfo {volume} {109}},\ \bibinfo {pages}
  {50001} (\bibinfo {year} {2015})}\BibitemShut {NoStop}%
\bibitem [{\citenamefont {Bhatnagar}\ \emph {et~al.}(1954)\citenamefont
  {Bhatnagar}, \citenamefont {Gross},\ and\ \citenamefont {Krook}}]{bgk1954}%
  \BibitemOpen
  \bibfield  {author} {\bibinfo {author} {\bibfnamefont {P.~L.}\ \bibnamefont
  {Bhatnagar}}, \bibinfo {author} {\bibfnamefont {E.~P.}\ \bibnamefont
  {Gross}}, \ and\ \bibinfo {author} {\bibfnamefont {M.}~\bibnamefont
  {Krook}},\ }\href@noop {} {\bibfield  {journal} {\bibinfo  {journal}
  {Physical Review}\ }\textbf {\bibinfo {volume} {94}} (\bibinfo {year}
  {1954})}\BibitemShut {NoStop}%
\bibitem [{\citenamefont {Dellar}(2001)}]{dellar2001bulk}%
  \BibitemOpen
  \bibfield  {author} {\bibinfo {author} {\bibfnamefont {P.~J.}\ \bibnamefont
  {Dellar}},\ }\href@noop {} {\bibfield  {journal} {\bibinfo  {journal}
  {Physical Review E}\ }\textbf {\bibinfo {volume} {64}},\ \bibinfo {pages}
  {031203} (\bibinfo {year} {2001})}\BibitemShut {NoStop}%
\bibitem [{\citenamefont {d'Humi{\`e}res}(2002)}]{d2002multiple}%
  \BibitemOpen
  \bibfield  {author} {\bibinfo {author} {\bibfnamefont {D.}~\bibnamefont
  {d'Humi{\`e}res}},\ }\href@noop {} {\bibfield  {journal} {\bibinfo  {journal}
  {Philosophical Transactions of the Royal Society of London A: Mathematical,
  Physical and Engineering Sciences}\ }\textbf {\bibinfo {volume} {360}},\
  \bibinfo {pages} {437} (\bibinfo {year} {2002})}\BibitemShut {NoStop}%
\bibitem [{\citenamefont {Latt}\ and\ \citenamefont
  {Chopard}(2006)}]{latt2006lattice}%
  \BibitemOpen
  \bibfield  {author} {\bibinfo {author} {\bibfnamefont {J.}~\bibnamefont
  {Latt}}\ and\ \bibinfo {author} {\bibfnamefont {B.}~\bibnamefont {Chopard}},\
  }\href@noop {} {\bibfield  {journal} {\bibinfo  {journal} {Mathematics and
  Computers in Simulation}\ }\textbf {\bibinfo {volume} {72}},\ \bibinfo
  {pages} {165} (\bibinfo {year} {2006})}\BibitemShut {NoStop}%
\bibitem [{\citenamefont {Zhang}\ \emph {et~al.}(2006)\citenamefont {Zhang},
  \citenamefont {Shan},\ and\ \citenamefont {Chen}}]{zhang2006efficient}%
  \BibitemOpen
  \bibfield  {author} {\bibinfo {author} {\bibfnamefont {R.}~\bibnamefont
  {Zhang}}, \bibinfo {author} {\bibfnamefont {X.}~\bibnamefont {Shan}}, \ and\
  \bibinfo {author} {\bibfnamefont {H.}~\bibnamefont {Chen}},\ }\href@noop {}
  {\bibfield  {journal} {\bibinfo  {journal} {Physical Review E}\ }\textbf
  {\bibinfo {volume} {74}},\ \bibinfo {pages} {046703} (\bibinfo {year}
  {2006})}\BibitemShut {NoStop}%
\bibitem [{\citenamefont {Karlin}\ \emph {et~al.}(1999)\citenamefont {Karlin},
  \citenamefont {Ferrante},\ and\ \citenamefont {{\"{O}}ttinger}}]{Karlin1999}%
  \BibitemOpen
  \bibfield  {author} {\bibinfo {author} {\bibfnamefont {I.~V.}\ \bibnamefont
  {Karlin}}, \bibinfo {author} {\bibfnamefont {A.}~\bibnamefont {Ferrante}}, \
  and\ \bibinfo {author} {\bibfnamefont {H.~C.}\ \bibnamefont
  {{\"{O}}ttinger}},\ }\href {\doibase 10.1209/epl/i1999-00370-1} {\bibfield
  {journal} {\bibinfo  {journal} {Europhysics Letters}\ }\textbf {\bibinfo
  {volume} {47}},\ \bibinfo {pages} {182} (\bibinfo {year} {1999})}\BibitemShut
  {NoStop}%
\bibitem [{\citenamefont {Karlin}\ \emph {et~al.}(2014)\citenamefont {Karlin},
  \citenamefont {B{\"{o}}sch},\ and\ \citenamefont
  {Chikatamarla}}]{Karlin2014}%
  \BibitemOpen
  \bibfield  {author} {\bibinfo {author} {\bibfnamefont {I.~V.}\ \bibnamefont
  {Karlin}}, \bibinfo {author} {\bibfnamefont {F.}~\bibnamefont {B{\"{o}}sch}},
  \ and\ \bibinfo {author} {\bibfnamefont {S.~S.}\ \bibnamefont
  {Chikatamarla}},\ }\href {\doibase 10.1103/PhysRevE.90.031302} {\bibfield
  {journal} {\bibinfo  {journal} {Physical Review E}\ }\textbf {\bibinfo
  {volume} {90}},\ \bibinfo {pages} {31302} (\bibinfo {year}
  {2014})}\BibitemShut {NoStop}%
\bibitem [{\citenamefont {Chikatamarla}\ and\ \citenamefont
  {Karlin}(2013)}]{Chikatamarla2013}%
  \BibitemOpen
  \bibfield  {author} {\bibinfo {author} {\bibfnamefont {S.}~\bibnamefont
  {Chikatamarla}}\ and\ \bibinfo {author} {\bibfnamefont {I.~V.}\ \bibnamefont
  {Karlin}},\ }\href {\doibase 10.1016/j.physa.2012.12.034} {\bibfield
  {journal} {\bibinfo  {journal} {Physica A}\ }\textbf {\bibinfo {volume}
  {392}},\ \bibinfo {pages} {1925} (\bibinfo {year} {2013})}\BibitemShut
  {NoStop}%
\bibitem [{\citenamefont {B{\"{o}}sch}\ \emph {et~al.}(2015)\citenamefont
  {B{\"{o}}sch}, \citenamefont {Chikatamarla},\ and\ \citenamefont
  {Karlin}}]{Bosch2015}%
  \BibitemOpen
  \bibfield  {author} {\bibinfo {author} {\bibfnamefont {F.}~\bibnamefont
  {B{\"{o}}sch}}, \bibinfo {author} {\bibfnamefont {S.~S.}\ \bibnamefont
  {Chikatamarla}}, \ and\ \bibinfo {author} {\bibfnamefont {I.~V.}\
  \bibnamefont {Karlin}},\ }\href {\doibase 10.1103/PhysRevE.92.043309}
  {\bibfield  {journal} {\bibinfo  {journal} {Physical Review E}\ }\textbf
  {\bibinfo {volume} {043309}},\ \bibinfo {pages} {1} (\bibinfo {year}
  {2015})}\BibitemShut {NoStop}%
\bibitem [{\citenamefont {Dorschner}\ \emph {et~al.}(2016)\citenamefont
  {Dorschner}, \citenamefont {Bösch}, \citenamefont {Chikatamarla},
  \citenamefont {Boulouchos},\ and\ \citenamefont {Karlin}}]{Dorschner2016}%
  \BibitemOpen
  \bibfield  {author} {\bibinfo {author} {\bibfnamefont {B.}~\bibnamefont
  {Dorschner}}, \bibinfo {author} {\bibfnamefont {F.}~\bibnamefont {Bösch}},
  \bibinfo {author} {\bibfnamefont {S.~S.}\ \bibnamefont {Chikatamarla}},
  \bibinfo {author} {\bibfnamefont {K.}~\bibnamefont {Boulouchos}}, \ and\
  \bibinfo {author} {\bibfnamefont {I.~V.}\ \bibnamefont {Karlin}},\ }\href
  {\doibase 10.1017/jfm.2016.448} {\bibfield  {journal} {\bibinfo  {journal}
  {Journal of Fluid Mechanics}\ }\textbf {\bibinfo {volume} {801}},\ \bibinfo
  {pages} {623} (\bibinfo {year} {2016})}\BibitemShut {NoStop}%
\bibitem [{\citenamefont {Ansumali}\ and\ \citenamefont
  {Karlin}(2005)}]{Ansumali2005b}%
  \BibitemOpen
  \bibfield  {author} {\bibinfo {author} {\bibfnamefont {S.}~\bibnamefont
  {Ansumali}}\ and\ \bibinfo {author} {\bibfnamefont {I.~V.}\ \bibnamefont
  {Karlin}},\ }\href {\doibase 10.1103/PhysRevLett.95.260605} {\bibfield
  {journal} {\bibinfo  {journal} {Physical Review Letters}\ }\textbf {\bibinfo
  {volume} {95}},\ \bibinfo {pages} {1} (\bibinfo {year} {2005})}\BibitemShut
  {NoStop}%
\bibitem [{\citenamefont {Chikatamarla}\ and\ \citenamefont
  {Karlin}(2009)}]{Chikatamarla2009}%
  \BibitemOpen
  \bibfield  {author} {\bibinfo {author} {\bibfnamefont {S.~S.}\ \bibnamefont
  {Chikatamarla}}\ and\ \bibinfo {author} {\bibfnamefont {I.~V.}\ \bibnamefont
  {Karlin}},\ }\href {\doibase 10.1103/PhysRevE.79.046701} {\bibfield
  {journal} {\bibinfo  {journal} {Physical Review E}\ }\textbf {\bibinfo
  {volume} {79}},\ \bibinfo {pages} {046701} (\bibinfo {year}
  {2009})}\BibitemShut {NoStop}%
\bibitem [{\citenamefont {Frapolli}\ \emph {et~al.}(2014)\citenamefont
  {Frapolli}, \citenamefont {Chikatamarla},\ and\ \citenamefont
  {Karlin}}]{frapolli2014multispeed}%
  \BibitemOpen
  \bibfield  {author} {\bibinfo {author} {\bibfnamefont {N.}~\bibnamefont
  {Frapolli}}, \bibinfo {author} {\bibfnamefont {S.~S.}\ \bibnamefont
  {Chikatamarla}}, \ and\ \bibinfo {author} {\bibfnamefont {I.~V.}\
  \bibnamefont {Karlin}},\ }\href@noop {} {\bibfield  {journal} {\bibinfo
  {journal} {Physical Review E}\ }\textbf {\bibinfo {volume} {90}},\ \bibinfo
  {pages} {043306} (\bibinfo {year} {2014})}\BibitemShut {NoStop}%
\bibitem [{\citenamefont {Frapolli}\ \emph {et~al.}(2015)\citenamefont
  {Frapolli}, \citenamefont {Chikatamarla},\ and\ \citenamefont
  {Karlin}}]{Frapolli2015}%
  \BibitemOpen
  \bibfield  {author} {\bibinfo {author} {\bibfnamefont {N.}~\bibnamefont
  {Frapolli}}, \bibinfo {author} {\bibfnamefont {S.~S.}\ \bibnamefont
  {Chikatamarla}}, \ and\ \bibinfo {author} {\bibfnamefont {I.~V.}\
  \bibnamefont {Karlin}},\ }\href {\doibase 10.1103/PhysRevE.92.061301}
  {\bibfield  {journal} {\bibinfo  {journal} {Physical Review E}\ }\textbf
  {\bibinfo {volume} {92}},\ \bibinfo {pages} {061301} (\bibinfo {year}
  {2015})}\BibitemShut {NoStop}%
\bibitem [{\citenamefont {Chen}\ \emph {et~al.}(2006)\citenamefont {Chen},
  \citenamefont {Filippova}, \citenamefont {Hoch}, \citenamefont {Molvig},
  \citenamefont {Shock}, \citenamefont {Teixeira},\ and\ \citenamefont
  {Zhang}}]{Chen2006}%
  \BibitemOpen
  \bibfield  {author} {\bibinfo {author} {\bibfnamefont {H.}~\bibnamefont
  {Chen}}, \bibinfo {author} {\bibfnamefont {O.}~\bibnamefont {Filippova}},
  \bibinfo {author} {\bibfnamefont {J.}~\bibnamefont {Hoch}}, \bibinfo {author}
  {\bibfnamefont {K.}~\bibnamefont {Molvig}}, \bibinfo {author} {\bibfnamefont
  {R.}~\bibnamefont {Shock}}, \bibinfo {author} {\bibfnamefont
  {C.}~\bibnamefont {Teixeira}}, \ and\ \bibinfo {author} {\bibfnamefont
  {R.}~\bibnamefont {Zhang}},\ }\href {\doibase 10.1016/j.physa.2005.09.036}
  {\bibfield  {journal} {\bibinfo  {journal} {Physica A: Statistical Mechanics
  and its Applications}\ }\textbf {\bibinfo {volume} {362}},\ \bibinfo {pages}
  {158} (\bibinfo {year} {2006})}\BibitemShut {NoStop}%
\bibitem [{\citenamefont {Rohde}\ \emph {et~al.}(2006)\citenamefont {Rohde},
  \citenamefont {Kandhai}, \citenamefont {Derksen},\ and\ \citenamefont
  {van~den Akker}}]{Rohde2006}%
  \BibitemOpen
  \bibfield  {author} {\bibinfo {author} {\bibfnamefont {M.}~\bibnamefont
  {Rohde}}, \bibinfo {author} {\bibfnamefont {D.}~\bibnamefont {Kandhai}},
  \bibinfo {author} {\bibfnamefont {J.~J.}\ \bibnamefont {Derksen}}, \ and\
  \bibinfo {author} {\bibfnamefont {H.~E.~A.}\ \bibnamefont {van~den Akker}},\
  }\href {\doibase 10.1002/fld.1140} {\bibfield  {journal} {\bibinfo  {journal}
  {International Journal for Numerical Methods in Fluids}\ }\textbf {\bibinfo
  {volume} {51}},\ \bibinfo {pages} {439} (\bibinfo {year} {2006})}\BibitemShut
  {NoStop}%
\bibitem [{\citenamefont {Filippova}\ and\ \citenamefont
  {H{\"{a}}nel}(1998)}]{Filippova1998}%
  \BibitemOpen
  \bibfield  {author} {\bibinfo {author} {\bibfnamefont {O.}~\bibnamefont
  {Filippova}}\ and\ \bibinfo {author} {\bibfnamefont {D.}~\bibnamefont
  {H{\"{a}}nel}},\ }\href {\doibase 10.1142/S012918319800114X} {\bibfield
  {journal} {\bibinfo  {journal} {International Journal of Modern Physics C}\
  }\textbf {\bibinfo {volume} {09}},\ \bibinfo {pages} {1271} (\bibinfo {year}
  {1998})}\BibitemShut {NoStop}%
\bibitem [{\citenamefont {Dupuis}\ and\ \citenamefont
  {Chopard}(2003)}]{Dupuis2003}%
  \BibitemOpen
  \bibfield  {author} {\bibinfo {author} {\bibfnamefont {A.}~\bibnamefont
  {Dupuis}}\ and\ \bibinfo {author} {\bibfnamefont {B.}~\bibnamefont
  {Chopard}},\ }\href {\doibase 10.1103/PhysRevE.67.066707} {\bibfield
  {journal} {\bibinfo  {journal} {Physical Review E}\ }\textbf {\bibinfo
  {volume} {67}},\ \bibinfo {pages} {066707} (\bibinfo {year}
  {2003})}\BibitemShut {NoStop}%
\bibitem [{\citenamefont {T{\"{o}}lke}\ and\ \citenamefont
  {Krafczyk}(2009)}]{Tolke2009}%
  \BibitemOpen
  \bibfield  {author} {\bibinfo {author} {\bibfnamefont {J.}~\bibnamefont
  {T{\"{o}}lke}}\ and\ \bibinfo {author} {\bibfnamefont {M.}~\bibnamefont
  {Krafczyk}},\ }\href {\doibase 10.1016/j.camwa.2009.02.012} {\bibfield
  {journal} {\bibinfo  {journal} {Computers {\&} Mathematics with
  Applications}\ }\textbf {\bibinfo {volume} {58}},\ \bibinfo {pages} {898}
  (\bibinfo {year} {2009})}\BibitemShut {NoStop}%
\bibitem [{\citenamefont {Lagrava}\ \emph {et~al.}(2012)\citenamefont
  {Lagrava}, \citenamefont {Malaspinas}, \citenamefont {Latt},\ and\
  \citenamefont {Chopard}}]{Lagrava2012}%
  \BibitemOpen
  \bibfield  {author} {\bibinfo {author} {\bibfnamefont {D.}~\bibnamefont
  {Lagrava}}, \bibinfo {author} {\bibfnamefont {O.}~\bibnamefont {Malaspinas}},
  \bibinfo {author} {\bibfnamefont {J.}~\bibnamefont {Latt}}, \ and\ \bibinfo
  {author} {\bibfnamefont {B.}~\bibnamefont {Chopard}},\ }\href {\doibase
  10.1016/j.jcp.2012.03.015} {\bibfield  {journal} {\bibinfo  {journal}
  {Journal of Computational Physics}\ }\textbf {\bibinfo {volume} {231}},\
  \bibinfo {pages} {4808} (\bibinfo {year} {2012})}\BibitemShut {NoStop}%
\bibitem [{\citenamefont {Ansumali}\ \emph {et~al.}(2003)\citenamefont
  {Ansumali}, \citenamefont {Karlin},\ and\ \citenamefont
  {{\"{O}}ttinger}}]{Ansumali2003a}%
  \BibitemOpen
  \bibfield  {author} {\bibinfo {author} {\bibfnamefont {S.}~\bibnamefont
  {Ansumali}}, \bibinfo {author} {\bibfnamefont {I.~V.}\ \bibnamefont
  {Karlin}}, \ and\ \bibinfo {author} {\bibfnamefont {H.~C.}\ \bibnamefont
  {{\"{O}}ttinger}},\ }\href {\doibase 10.1209/epl/i2003-00496-6} {\bibfield
  {journal} {\bibinfo  {journal} {Europhysics Letters}\ }\textbf {\bibinfo
  {volume} {63}},\ \bibinfo {pages} {798} (\bibinfo {year} {2003})}\BibitemShut
  {NoStop}%
\bibitem [{\citenamefont {Ansumali}\ \emph {et~al.}(2007)\citenamefont
  {Ansumali}, \citenamefont {Arcidiacono}, \citenamefont {Chikatamarla},
  \citenamefont {Prasianakis}, \citenamefont {Gorban},\ and\ \citenamefont
  {Karlin}}]{ansumali2007quasi}%
  \BibitemOpen
  \bibfield  {author} {\bibinfo {author} {\bibfnamefont {S.}~\bibnamefont
  {Ansumali}}, \bibinfo {author} {\bibfnamefont {S.}~\bibnamefont
  {Arcidiacono}}, \bibinfo {author} {\bibfnamefont {S.}~\bibnamefont
  {Chikatamarla}}, \bibinfo {author} {\bibfnamefont {N.}~\bibnamefont
  {Prasianakis}}, \bibinfo {author} {\bibfnamefont {A.}~\bibnamefont {Gorban}},
  \ and\ \bibinfo {author} {\bibfnamefont {I.}~\bibnamefont {Karlin}},\
  }\href@noop {} {\bibfield  {journal} {\bibinfo  {journal} {The European
  Physical Journal B}\ }\textbf {\bibinfo {volume} {56}},\ \bibinfo {pages}
  {135} (\bibinfo {year} {2007})}\BibitemShut {NoStop}%
\bibitem [{\citenamefont {Karlin}\ \emph {et~al.}(2013)\citenamefont {Karlin},
  \citenamefont {Sichau},\ and\ \citenamefont {Chikatamarla}}]{Karlin2013}%
  \BibitemOpen
  \bibfield  {author} {\bibinfo {author} {\bibfnamefont {I.~V.}\ \bibnamefont
  {Karlin}}, \bibinfo {author} {\bibfnamefont {D.}~\bibnamefont {Sichau}}, \
  and\ \bibinfo {author} {\bibfnamefont {S.~S.}\ \bibnamefont {Chikatamarla}},\
  }\href {\doibase 10.1103/PhysRevE.88.063310} {\bibfield  {journal} {\bibinfo
  {journal} {Physical Review E}\ }\textbf {\bibinfo {volume} {88}},\ \bibinfo
  {pages} {1} (\bibinfo {year} {2013})}\BibitemShut {NoStop}%
\bibitem [{\citenamefont {Pareschi}\ \emph {et~al.}(2016)\citenamefont
  {Pareschi}, \citenamefont {Frapolli}, \citenamefont {Chikatamarla},\ and\
  \citenamefont {Karlin}}]{pareschi2016thermal}%
  \BibitemOpen
  \bibfield  {author} {\bibinfo {author} {\bibfnamefont {G.}~\bibnamefont
  {Pareschi}}, \bibinfo {author} {\bibfnamefont {N.}~\bibnamefont {Frapolli}},
  \bibinfo {author} {\bibfnamefont {S.}~\bibnamefont {Chikatamarla}}, \ and\
  \bibinfo {author} {\bibfnamefont {I.}~\bibnamefont {Karlin}},\ }\href@noop {}
  {\bibfield  {journal} {\bibinfo  {journal} {Physical Review E}\ }\textbf
  {\bibinfo {volume} {94}},\ \bibinfo {pages} {013305} (\bibinfo {year}
  {2016})}\BibitemShut {NoStop}%
\bibitem [{\citenamefont {Frapolli}\ \emph
  {et~al.}(2016{\natexlab{a}})\citenamefont {Frapolli}, \citenamefont
  {Chikatamarla},\ and\ \citenamefont {Karlin}}]{frapolli2016compressible}%
  \BibitemOpen
  \bibfield  {author} {\bibinfo {author} {\bibfnamefont {N.}~\bibnamefont
  {Frapolli}}, \bibinfo {author} {\bibfnamefont {S.}~\bibnamefont
  {Chikatamarla}}, \ and\ \bibinfo {author} {\bibfnamefont {I.}~\bibnamefont
  {Karlin}},\ }\href@noop {} {\bibfield  {journal} {\bibinfo  {journal}
  {Physical Review E}\ }\textbf {\bibinfo {volume} {93}},\ \bibinfo {pages}
  {063302} (\bibinfo {year} {2016}{\natexlab{a}})}\BibitemShut {NoStop}%
\bibitem [{\citenamefont {Kupershtokh}(2004)}]{Kupershtokh2004}%
  \BibitemOpen
  \bibfield  {author} {\bibinfo {author} {\bibfnamefont {A.~L.}\ \bibnamefont
  {Kupershtokh}},\ }in\ \href@noop {} {\emph {\bibinfo {booktitle} {Proc. 5th
  International EHD Workshop}}}\ (\bibinfo {year} {2004})\ pp.\ \bibinfo
  {pages} {241--246}\BibitemShut {NoStop}%
\bibitem [{\citenamefont {Dorschner}\ \emph {et~al.}(2015)\citenamefont
  {Dorschner}, \citenamefont {Chikatamarla}, \citenamefont {B{\"{o}}sch},\ and\
  \citenamefont {Karlin}}]{Dorschner2015}%
  \BibitemOpen
  \bibfield  {author} {\bibinfo {author} {\bibfnamefont {B.}~\bibnamefont
  {Dorschner}}, \bibinfo {author} {\bibfnamefont {S.}~\bibnamefont
  {Chikatamarla}}, \bibinfo {author} {\bibfnamefont {F.}~\bibnamefont
  {B{\"{o}}sch}}, \ and\ \bibinfo {author} {\bibfnamefont {I.}~\bibnamefont
  {Karlin}},\ }\href {\doibase 10.1016/j.jcp.2015.04.017} {\bibfield  {journal}
  {\bibinfo  {journal} {Journal of Computational Physics}\ }\textbf {\bibinfo
  {volume} {295}},\ \bibinfo {pages} {340} (\bibinfo {year}
  {2015})}\BibitemShut {NoStop}%
\bibitem [{\citenamefont {Moser}\ \emph {et~al.}(1999)\citenamefont {Moser},
  \citenamefont {Kim},\ and\ \citenamefont {Mansour}}]{Moser1999}%
  \BibitemOpen
  \bibfield  {author} {\bibinfo {author} {\bibfnamefont {R.~D.}\ \bibnamefont
  {Moser}}, \bibinfo {author} {\bibfnamefont {J.}~\bibnamefont {Kim}}, \ and\
  \bibinfo {author} {\bibfnamefont {N.~N.}\ \bibnamefont {Mansour}},\ }\href
  {\doibase 10.1063/1.869966} {\bibfield  {journal} {\bibinfo  {journal}
  {Physics of Fluids}\ }\textbf {\bibinfo {volume} {11}},\ \bibinfo {pages}
  {943} (\bibinfo {year} {1999})}\BibitemShut {NoStop}%
\bibitem [{\citenamefont {Eckelmann}(1974)}]{Eckelmann1974}%
  \BibitemOpen
  \bibfield  {author} {\bibinfo {author} {\bibfnamefont {H.}~\bibnamefont
  {Eckelmann}},\ }\href {\doibase 10.1017/S0022112074001479} {\bibfield
  {journal} {\bibinfo  {journal} {Journal of Fluid Mechanics}\ }\textbf
  {\bibinfo {volume} {65}},\ \bibinfo {pages} {439} (\bibinfo {year}
  {1974})}\BibitemShut {NoStop}%
\bibitem [{\citenamefont {Kreplin}\ and\ \citenamefont
  {Eckelmann}(1979)}]{Kreplin1979}%
  \BibitemOpen
  \bibfield  {author} {\bibinfo {author} {\bibfnamefont {H.-P.}\ \bibnamefont
  {Kreplin}}\ and\ \bibinfo {author} {\bibfnamefont {H.}~\bibnamefont
  {Eckelmann}},\ }\href {\doibase 10.1063/1.862737} {\bibfield  {journal}
  {\bibinfo  {journal} {Physics of Fluids}\ }\textbf {\bibinfo {volume} {22}},\
  \bibinfo {pages} {1233} (\bibinfo {year} {1979})}\BibitemShut {NoStop}%
\bibitem [{\citenamefont {Kim}\ \emph {et~al.}(1987)\citenamefont {Kim},
  \citenamefont {Moin},\ and\ \citenamefont {Moser}}]{Kim1987}%
  \BibitemOpen
  \bibfield  {author} {\bibinfo {author} {\bibfnamefont {J.}~\bibnamefont
  {Kim}}, \bibinfo {author} {\bibfnamefont {P.}~\bibnamefont {Moin}}, \ and\
  \bibinfo {author} {\bibfnamefont {R.}~\bibnamefont {Moser}},\ }\href
  {\doibase 10.1017/S0022112087000892} {\bibfield  {journal} {\bibinfo
  {journal} {Journal of Fluid Mechanics}\ }\textbf {\bibinfo {volume} {177}},\
  \bibinfo {pages} {133} (\bibinfo {year} {1987})}\BibitemShut {NoStop}%
\bibitem [{\citenamefont {Pope}(2000)}]{pope2000turbulent}%
  \BibitemOpen
  \bibfield  {author} {\bibinfo {author} {\bibfnamefont {S.~B.}\ \bibnamefont
  {Pope}},\ }\href@noop {} {\emph {\bibinfo {title} {{Turbulent flows}}}}\
  (\bibinfo  {publisher} {Cambridge university press},\ \bibinfo {year}
  {2000})\BibitemShut {NoStop}%
\bibitem [{\citenamefont {Rodriguez}\ \emph {et~al.}(2011)\citenamefont
  {Rodriguez}, \citenamefont {Borell}, \citenamefont {Lehmkuhl}, \citenamefont
  {{Perez Segarra}},\ and\ \citenamefont {Oliva}}]{Rodriguez2011}%
  \BibitemOpen
  \bibfield  {author} {\bibinfo {author} {\bibfnamefont {I.}~\bibnamefont
  {Rodriguez}}, \bibinfo {author} {\bibfnamefont {R.}~\bibnamefont {Borell}},
  \bibinfo {author} {\bibfnamefont {O.}~\bibnamefont {Lehmkuhl}}, \bibinfo
  {author} {\bibfnamefont {C.~D.}\ \bibnamefont {{Perez Segarra}}}, \ and\
  \bibinfo {author} {\bibfnamefont {A.}~\bibnamefont {Oliva}},\ }\href
  {\doibase 10.1017/jfm.2011.136} {\bibfield  {journal} {\bibinfo  {journal}
  {Journal of Fluid Mechanics}\ }\textbf {\bibinfo {volume} {679}},\ \bibinfo
  {pages} {263} (\bibinfo {year} {2011})}\BibitemShut {NoStop}%
\bibitem [{\citenamefont {Yun}\ \emph {et~al.}(2006)\citenamefont {Yun},
  \citenamefont {Kim},\ and\ \citenamefont {Choi}}]{Yun2006}%
  \BibitemOpen
  \bibfield  {author} {\bibinfo {author} {\bibfnamefont {G.}~\bibnamefont
  {Yun}}, \bibinfo {author} {\bibfnamefont {D.}~\bibnamefont {Kim}}, \ and\
  \bibinfo {author} {\bibfnamefont {H.}~\bibnamefont {Choi}},\ }\href {\doibase
  10.1063/1.2166454} {\bibfield  {journal} {\bibinfo  {journal} {Physics of
  Fluids}\ }\textbf {\bibinfo {volume} {18}} (\bibinfo {year} {2006}),\
  10.1063/1.2166454}\BibitemShut {NoStop}%
\bibitem [{\citenamefont {{Kim, H.J., Durbin}}(1988)}]{KimH.J.Durbin1988}%
  \BibitemOpen
  \bibfield  {author} {\bibinfo {author} {\bibfnamefont {P.}~\bibnamefont
  {{Kim, H.J., Durbin}}},\ }\href {\doibase 10.1063/1.866937} {\bibfield
  {journal} {\bibinfo  {journal} {Physics of Fluids}\ }\textbf {\bibinfo
  {volume} {31}},\ \bibinfo {pages} {3260} (\bibinfo {year}
  {1988})}\BibitemShut {NoStop}%
\bibitem [{\citenamefont {Schlichting}\ and\ \citenamefont
  {Gersten}(2003)}]{schlichting2003}%
  \BibitemOpen
  \bibfield  {author} {\bibinfo {author} {\bibfnamefont {H.}~\bibnamefont
  {Schlichting}}\ and\ \bibinfo {author} {\bibfnamefont {K.}~\bibnamefont
  {Gersten}},\ }\href@noop {} {\emph {\bibinfo {title} {{Boundary-layer
  theory}}}}\ (\bibinfo  {publisher} {Springer Science {\&} Business Media},\
  \bibinfo {year} {2003})\BibitemShut {NoStop}%
\bibitem [{\citenamefont {Togni}\ \emph {et~al.}(2015)\citenamefont {Togni},
  \citenamefont {Cimarelli},\ and\ \citenamefont
  {De~Angelis}}]{togni2015physical}%
  \BibitemOpen
  \bibfield  {author} {\bibinfo {author} {\bibfnamefont {R.}~\bibnamefont
  {Togni}}, \bibinfo {author} {\bibfnamefont {A.}~\bibnamefont {Cimarelli}}, \
  and\ \bibinfo {author} {\bibfnamefont {E.}~\bibnamefont {De~Angelis}},\
  }\href@noop {} {\bibfield  {journal} {\bibinfo  {journal} {Journal of Fluid
  Mechanics}\ }\textbf {\bibinfo {volume} {782}},\ \bibinfo {pages} {380}
  (\bibinfo {year} {2015})}\BibitemShut {NoStop}%
\bibitem [{\citenamefont {Venezian}\ \emph {et~al.}(1962)\citenamefont
  {Venezian}, \citenamefont {Crespo},\ and\ \citenamefont
  {Sage}}]{venezian1962thermal}%
  \BibitemOpen
  \bibfield  {author} {\bibinfo {author} {\bibfnamefont {E.}~\bibnamefont
  {Venezian}}, \bibinfo {author} {\bibfnamefont {M.~J.}\ \bibnamefont
  {Crespo}}, \ and\ \bibinfo {author} {\bibfnamefont {B.}~\bibnamefont
  {Sage}},\ }\href@noop {} {\bibfield  {journal} {\bibinfo  {journal} {AIChE
  Journal}\ }\textbf {\bibinfo {volume} {8}},\ \bibinfo {pages} {383} (\bibinfo
  {year} {1962})}\BibitemShut {NoStop}%
\bibitem [{\citenamefont {Frapolli}\ \emph
  {et~al.}(2016{\natexlab{b}})\citenamefont {Frapolli}, \citenamefont
  {Chikatamarla},\ and\ \citenamefont {Karlin}}]{frapolli2016lattice}%
  \BibitemOpen
  \bibfield  {author} {\bibinfo {author} {\bibfnamefont {N.}~\bibnamefont
  {Frapolli}}, \bibinfo {author} {\bibfnamefont {S.}~\bibnamefont
  {Chikatamarla}}, \ and\ \bibinfo {author} {\bibfnamefont {I.}~\bibnamefont
  {Karlin}},\ }\href@noop {} {\bibfield  {journal} {\bibinfo  {journal}
  {Physical Review Letters}\ }\textbf {\bibinfo {volume} {117}},\ \bibinfo
  {pages} {010604} (\bibinfo {year} {2016}{\natexlab{b}})}\BibitemShut
  {NoStop}%
\bibitem [{\citenamefont {Hafez}\ and\ \citenamefont
  {Wahba}(2007)}]{hafez2007simulations}%
  \BibitemOpen
  \bibfield  {author} {\bibinfo {author} {\bibfnamefont {M.}~\bibnamefont
  {Hafez}}\ and\ \bibinfo {author} {\bibfnamefont {E.}~\bibnamefont {Wahba}},\
  }\href@noop {} {\bibfield  {journal} {\bibinfo  {journal} {Computers \&
  Fluids}\ }\textbf {\bibinfo {volume} {36}},\ \bibinfo {pages} {39} (\bibinfo
  {year} {2007})}\BibitemShut {NoStop}%
\end{thebibliography}%

\end{document}